\documentclass[usenames,dvipsnames]{aastex62}

\usepackage{xcolor}

\usepackage{multirow}
\submitjournal{ApJ}

\shorttitle{}
\shortauthors{Zucker et al.}

\begin{document}

\title{Synthetic Large-Scale Galactic Filaments -- on their Formation, Physical Properties, and Resemblance to Observations}

\correspondingauthor{Catherine Zucker}
\author{Catherine Zucker}
\email{catherine.zucker@cfa.harvard.edu}
\affil{Harvard Astronomy, Harvard-Smithsonian Center for Astrophysics, 60 Garden St., Cambridge, MA 02138, USA}

\author{Rowan Smith}
\affil{Jodrell Bank Centre for Astrophysics, School of Physics and Astronomy, University of Manchester, Oxford Road,Manchester, M13 9PL,UK}

\author{Alyssa Goodman}
\affil{Harvard Astronomy, Harvard-Smithsonian Center for Astrophysics, 60 Garden St., Cambridge, MA 02138, USA}
\affil{Radcliffe Institute for Advanced Study, Harvard University, 10 Garden St, Cambridge, MA 02138}



\begin{abstract}
\noindent 
Using a population of large-scale filaments extracted from an AREPO simulation of a Milky Way-like galaxy, we seek to understand the extent to which observed large-scale filament properties (with lengths $\gtrsim 100$ pc) can be explained by galactic dynamics alone. From an observer's perspective in the disk of the galaxy, we identify filaments forming purely due to galactic dynamics, without the effects of feedback or local self-gravity. We find that large-scale Galactic filaments are intrinsically rare, and we estimate that at maximum approximately one filament per $\rm kpc^{2}$ should be identified in projection, when viewed from the direction of our Sun in the Milky Way.  In this idealized scenario, we find filaments in both the arm and interarm regions, and hypothesize that the former may be due to gas compression in the spiral-potential wells, with the latter due to differential rotation. Using the same analysis pipeline applied previously to observations, we analyze the physical properties of large-scale Galactic filaments, and quantify their sensitivity to projection effects and galactic environment (i.e. whether they lie in the arm or interarm regions). We find that observed ``Giant Molecular Filaments" are consistent with being non-self-gravitating structures dominated by galactic dynamics. Straighter, narrower, and denser ``Bone-like" filaments, like the paradigmatic Nessie filament, have similar column densities, velocity gradients, and Galactic plane heights ($z\approx$ 0 pc) to those in our simple model, but additional physical effects (such as feedback and self-gravity) must be invoked to explain their lengths and widths.
\end{abstract}

\keywords{}


\section{Introduction} \label{sec:intro}
As spiral arms are a prominent feature of Milky Way-like galaxies, it is crucial to understand how they affect the transformation of gas into stars within molecular clouds. We have known for decades \citep[e.g.][]{Roberts_1969} that molecular gas responds strongly to dynamical influences, and these dynamical influences govern the formation of structures on tens of parsecs to kiloparsec scales. For example, simulations have shown that the formation of spurs and feathers in spiral galaxies may be caused by shear arising from divergent orbits in the spiral potential, as dense molecular gas leaves the potential and is stretched out in the interarm regions \citep{Kim_2002, Dobbs_2006, Shetty_2006}. The arms themselves may also be critical to the formation of molecular clouds, with some models suggesting that spiral shocks induce smaller-scale, high-density structures in the arm, which agglomerate into molecular clouds \citep{Dobbs_2008, Bonnell_2006}. 

We now have resolved molecular cloud catalogs \citep{Rice_2016, Colombo_2019, Miville_Deschenes_2017} over the entire Galactic disc, with evidence that some molecular cloud properties (e.g. surface density, cloud brightness) are modestly higher in the spiral arms. A fraction of these molecular clouds have been shown to have abnormally high aspect ratios ($\approx $ 5:1 - 10:1). These so-called ``Giant Molecular Filaments" (or GMFs)  \citep{Ragan_2014, Abreu_Vicente_2016, Zhang_2019, Du_2017} appear to have masses and column densities similar to the typical molecular cloud, in spite of their atypical elongation. In tandem, other studies have cataloged much more elongated and dense molecular gas structures (the ``Bones" of the Milky Way) after the discovery by \citet{Goodman_2014} that the Nessie filament \citep{Jackson_2010} is even longer than originally claimed ($>$ 150 pc); aligns with the Scutum-Centaurus arm in \textit{position-position-velocity} space; and is likely formed and maintained by larger-scale Galactic forces due to its incredibly high ( $> 300:1$) aspect ratio. 

A recent study by \citet{Zucker_2018a} develops an observational pipeline to uniformly characterize the physical properties of all elongated molecular gas features purportedly associated with spiral structure, using the same datasets, statistical techniques, and spiral arm models \citep{Ragan_2014, Abreu_Vicente_2016, Wang_2015, Wang_2016, Zucker_2015}.  While most filaments are preferentially aligned and in close proximity spatially to the Galactic plane, \citet{Zucker_2018a} finds that kinematic association with purported spiral arm models is more rare, with less than half the sample displaying velocities consistent with spiral features. \citet{Zucker_2018a} also finds large catalog-to-catalog variations in filament properties, with some properties varying by an order of magnitude across the full sample. The filament's fraction of cold and dense gas, along with its aspect ratio, are able to broadly distinguish between different observed filament samples, and large variations in these properties could be indicative of different formation mechanisms or evolutionary histories. 

Due to their unique morphology, previous numerical studies suggest that the formation and evolution of these filaments may be tied to galactic dynamics, with different types of filaments (``Bones", "GMFs") potentially forming in different environments. \citet{Duarte_Cabral_2017}, for instance, uses a Smoothed Particle Hydrodynamics (SPH) simulation to follow two filaments in their evolution through the disc, finding that highly elongated filamentary structures are formed in the interarm regions via galactic shear, and only become largely molecular and aligned with spiral arms at the deepest point in the potential well, just prior to arm entry. Using the AREPO moving mesh code \citet{Smith_2014} studies a large swath of a spiral galaxy disc ($\rm \approx 30 \; kpc^{2}$) at a single time snapshot, reaching a resolution of 0.3 pc in regions of gas density greater than $\rm 10^{3} \;cm^{-3}$. Like \citet{Duarte_Cabral_2017}, \citet{Smith_2014} finds that highly elongated filaments, dominated by CO ``dark" gas, reside in long filaments stretched between spiral arms due to differential rotation, while the highest density filaments may form due to shocks in the spiral-potential wells. 

While both the \citet{Smith_2014} and \citet{Duarte_Cabral_2016, Duarte_Cabral_2017} simulations suggest that the physical properties of synthetic large-scale filaments may change due to galactic environment\footnote{Throughout this work we will use the term ``galactic environment" to denote variations in properties due to association with either the \textit{arm} or \textit{interarm} regions},  comparatively little work has been done to systematically analyze these properties and contextualize them in light of the growing sample of (very diverse) filaments observed in our own Galaxy \citep{Zucker_2018a}. The \citet{Smith_2014} simulations provide the opportunity to analyze the response of these dense molecular filaments to an external spiral potential at high resolution over a large fraction of the disc. In this work, we extract a sample of filaments from the perspective of an observer in the disc of the \citet{Smith_2014} simulation. To facilitate a direct comparison with observations, we analyze the properties of these filaments using the same methodology systematically applied to the observed large-scale filament population in \citet{Zucker_2018a}. This allows for a direct comparison between the synthetic and observed filament properties, to determine to what extent large-scale filaments properties can be explained by galactic dynamics alone. In \S \ref{sec:methods} we discuss our method for extracting grids from the \citet{Smith_2014} simulations, producing realistic $\rm H_2$ column density projections, identifying filaments, and post-processing them to obtain synthetic maps of molecular emission. In \S \ref{sec:results} we present the physical properties (length, width, column density, mass, line mass, galactic plane separation, position angle, velocity gradient) we compute for the sample. In \S \ref{sec:discussion}, we discuss the variation in these properties due projection effects and galactic environment (arm/interarm regions). We conclude in \S \ref{sec:conclusion}. 

\section{Methods} \label{sec:methods}
\subsection{Numerical Simulations} \label{subsec:sims}
Here we briefly describe the AREPO simulations \citep{Smith_2014} from which the filaments are extracted. For a complete overview of the model, see \citet[][\S 2.2-2.4]{Smith_2014}. AREPO \citep{Springel_2010} is a moving mesh code, where the unstructured mesh is defined by a Voronoi tesselation of discrete grid points that move with the flow. This flexibility of movement allows the mesh to smoothly adjust its spatial resolution, and also provides improved mass refinement in regions of interest. Its high dynamic range is ideal for study of large-scale galactic filaments, as it allows one to resolve both the narrow widths of these filaments and the environments in which they form. To produce a Milky Way analog, \citet{Smith_2014} adopts the Galactic potential of \citet{Dobbs_2006}, which is an analytic four-armed spiral potential imposed on a disc of gas. The simulations do {\it not} include local self-gravity or stellar feedback, but they do include a simple chemical model for CO and $\rm H_2$ chemistry, following \citet{Glover_2007a, Glover_2007b} and \citet{Nelson_1997}. This chemical network allows one to track the molecular hydrogen and CO abundances of the filaments, which are used to compute the $\rm H_2$ column density projections in \S \ref{subsec:projections} and the synthetic CO spectral cubes in \S \ref{subsec:radmc}. These simulations represent the minimum physics needed, as local self-gravity, stellar feedback, and magnetic fields may also play a role. Our setup allows us to identify when these are dominant effects and when they are secondary ones, which is discussed further in \S \ref{sec:results}. 

The initial condition of the simulation is a torus with a thickness of 200 pc, with an inner radius of 5 kpc and an outer radius of 10 kpc. The Galactic center region is excluded both for computational efficiency and because the gas dynamics in this region should be strongly influenced by the Galactic bar, of which there is strong evidence of in the Milky Way \citep[c.f.][]{Benjamin_2005}. The \citet{Dobbs_2006} potential causes the gas to move clockwise with a radial velocity of 220 $\rm km \; s^{-1}$. The gas is allowed to evolve over 1.5 rotations (260 Myr). In order to resolve individual molecular clouds, \citet{Smith_2014} takes advantage of AREPO's mass refinement scheme, and selects one section of the disc ($\rm \approx 30 \; kpc^{2}$) to increase the mass resolution to $\rm 4 \; M_\sun$. This is the section of the disc we utilize for region extraction in \S \ref{subsec:extraction}.

\subsection{Region Extraction} \label{subsec:extraction}
\begin{figure}[h!]
\begin{center}
\includegraphics[width=1.\columnwidth]{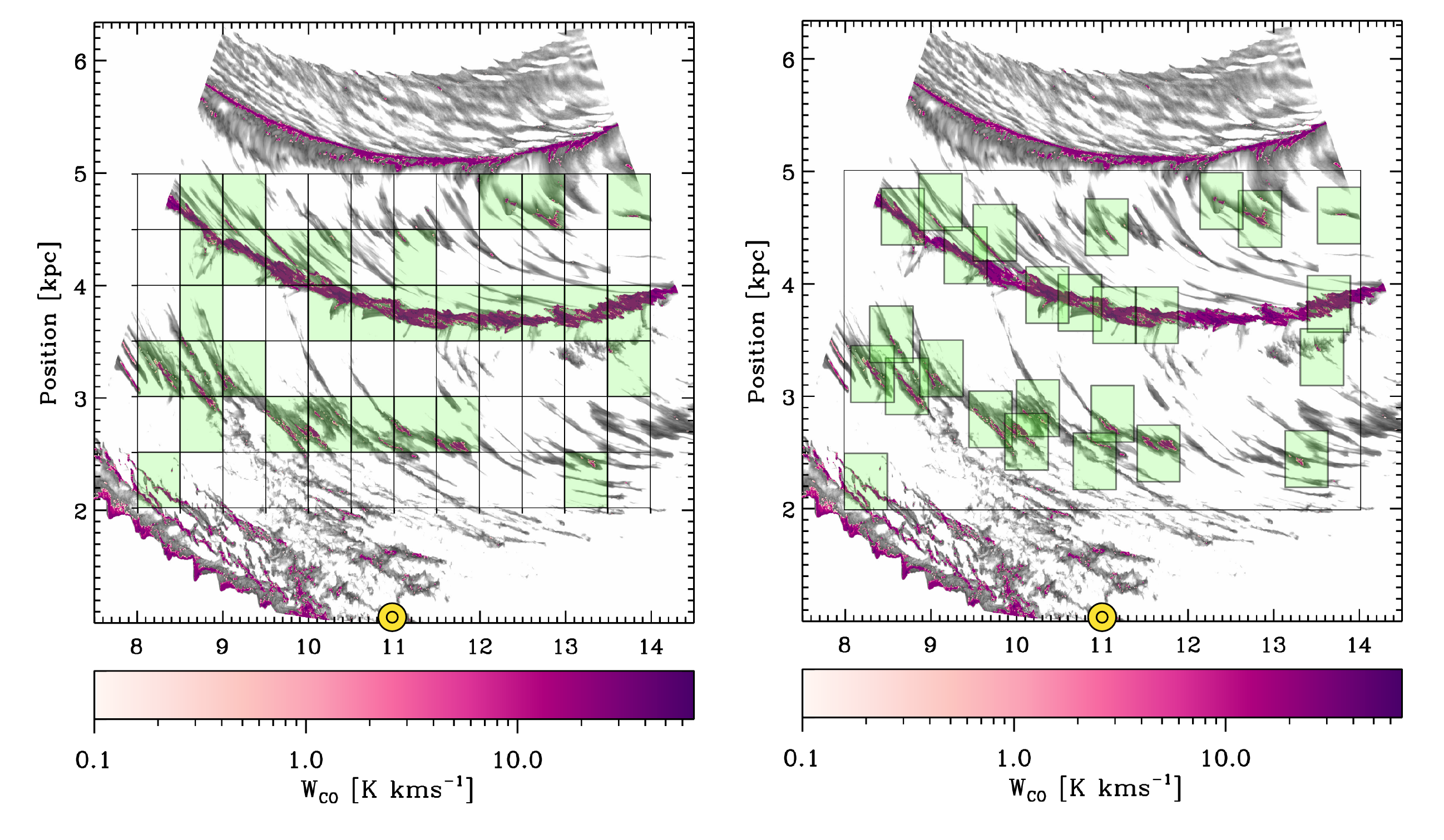}
\caption{ \label{fig:gridsearch} Top down CO integrated intensity map of a highly resolved section of the disc from \citet{Smith_2014}. \textbf{Left:} The disk is divided into 500 pc$^{3}$ boxes (black grid), and boxes containing extended (aspect ratio $\gtrsim3$) CO integrated intensity structures (above $\rm 1 \; K \; km \; s^{-1}$) are selected for further analysis (shown in light green). Our assumed ``Sun" position is marked with the yellow circle---(x,y)=(11 kpc, 1 kpc)---and is used in the determination of the ``sun-like" projections (see \S \ref{subsec:projections}). \textbf{Right:} After identifying filaments on the regular grids (left panel), we shift each grid so it is centered on the identified filament, to avoid gridding effects in the derived physical properties. The AREPO data for these grids are publicly available on the Harvard Dataverse (\href{https://doi.org/10.7910/DVN/NN0FLK}{doi:10.7910/DVN/NN0FLK})}
\end{center}
\end{figure}

To focus on regions of interest, we start by dividing the \citet{Smith_2014} ``zoom-in" --- the region of refined spatial and mass resolution ---  into boxes of size 500 pc$^3$. We consider the area between x=\{8 kpc, 14 kpc\}, y=\{2 kpc, 5 kpc\} in Figure \ref{fig:gridsearch}. We exclude the spiral arm closest to the Galactic center ($y > 5$ kpc), due to chemically immature gas streaming in from outside the zoom-in simulation at lower mass resolution. 

For computational expediency, we select only those grids with appreciable amounts of integrated CO emission (above $\rm 1 \; K\; km \; s^{-1}$) for further analysis, thereby excluding regions composed predominantly of CO-dark gas. This is consistent with spectral-line observations of large-scale filaments in the Milky Way, which all show continuous CO position-velocity tracks \citep{Zucker_2018a}. The 31 grids we select for the search are highlighted in green and overlaid on the highly resolved section of the disk in the left panel of Figure \ref{fig:gridsearch} (with the CO integrated intensity shown with a pink colorscale, and the $\rm H_2$ column density shown with a grayscale). The selection was done by eye, but as is apparent from Figure \ref{fig:gridsearch}, the remaining regions have no extended (aspect ratio $> 3:1$) structures above $\rm W(CO) \rm \approx 1 \; K \; km \; s^{-1}$, which precludes the possibility of them containing any dense gaseous filaments akin to those seen in CO observations \citep{Zucker_2015, Zucker_2018a, Wang_2015}. For each of the green boxes shown in the left panel of Figure \ref{fig:gridsearch}, we interpolate the AREPO mesh onto a fixed grid with a cell size of $\rm 1 \; pc^{3}$. The resolution obtained by Herschel in the $500 \micron$ band at the typical distance of observed large-scale galactic filaments \citep[3.3 kpc; see][]{Zucker_2018a} is 0.7 pc, which is on par with the regridded simulations. In order to avoid gridding effects (e.g. the filaments identified being split between two grids), we perform this process twice. That is, we first identify filaments inside the boxes shown in Figure \ref{fig:gridsearch}. Then, once filaments are identified, we shift each grid towards the central coordinate of each filament, so that is entirely contained with a single 500 pc$^{3}$ box and properties like length are unbiased by our choice of grid. The shifted grids are likewise shown in green in the right panel of Figure \ref{fig:gridsearch}. This procedure is discussed more in \S \ref{subsec:identification}. 

All the AREPO data used in this work (for the grids shown in the right panel of Figure \ref{fig:gridsearch}) are publicly available for download on the Harvard Dataverse (\href{https://doi.org/10.7910/DVN/NN0FLK}{doi:10.7910/DVN/NN0FLK}). 

\subsection{Projecting the Data} \label{subsec:projections}
\begin{figure}[h!]
\begin{center}
\includegraphics[width=1.\columnwidth]{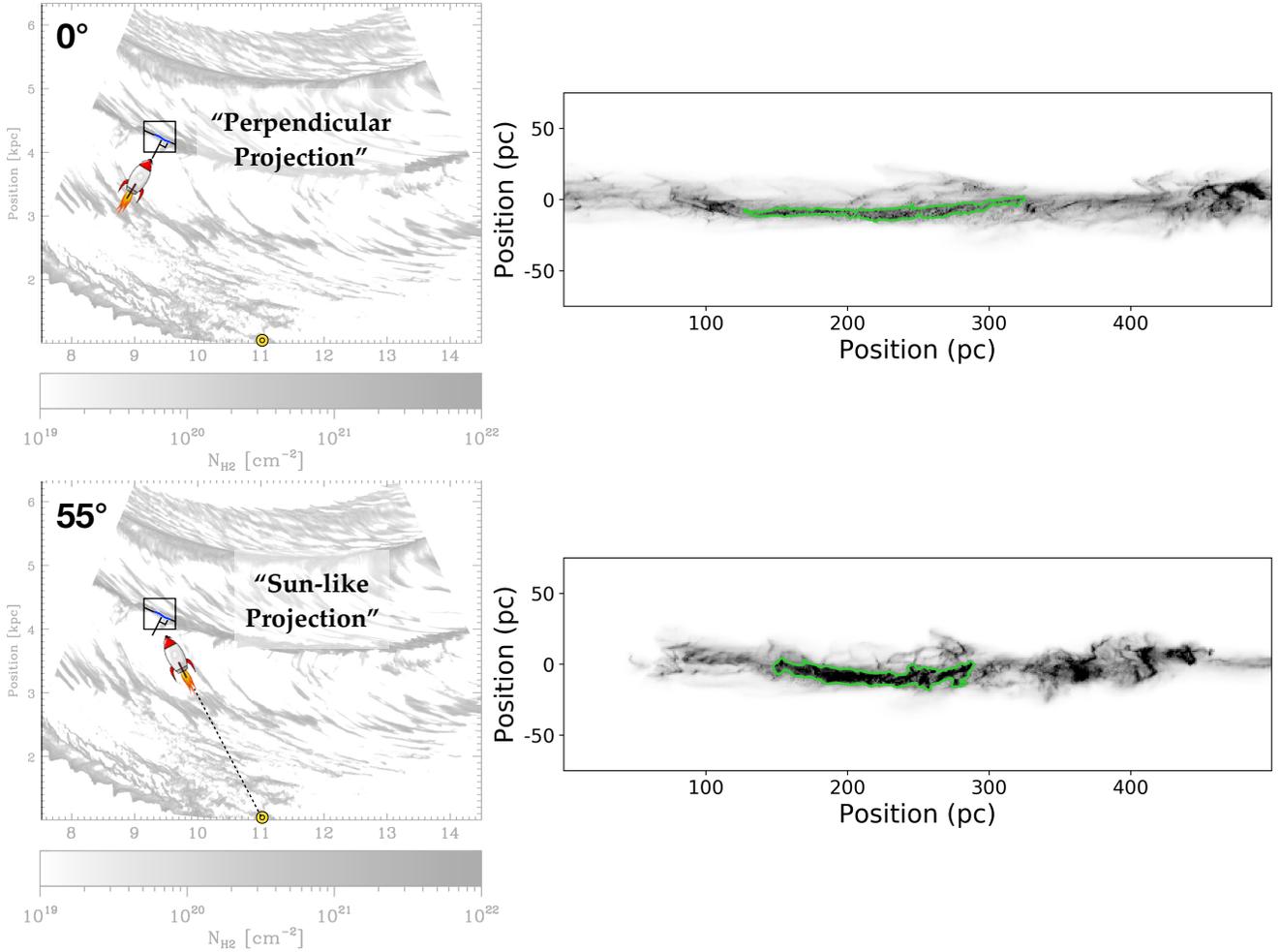}
\caption{ \label{fig:projections} Simple cartoon illustrating the two different projections we utilize. \textbf{Top:} In the ``perpendicular" projection, we assume that the vector pointing towards the observer (normal to the projection plane) is perpendicular to the long axis of the filament. This should mitigate the impact of projection effects on the physical properties we calculate. \textbf{Bottom:} In the ``sun-like" projection, the vector points towards an assumed sun position in the galaxy (x,y)=(11 kpc, 1 kpc), which is chosen to roughly mimic the configuration from which we observe the Scutum-Centaurus arm in the Milky Way, where most of the large-scale Galactic filament population currently resides. \textbf{For an animated version of this figure, see the YouTube video at \href{https://youtu.be/3qbYvreT2AQ}{https://youtu.be/3qbYvreT2AQ}}.}
\end{center}
\end{figure}

To produce realistic $\rm H_2$ column density maps of the extracted grids (described in \S \ref{subsec:extraction}) we project the data in two ways to identify filaments (hereafter the ``perpendicular" and ``sun-like" projections). The two projections are illustrated in Figure \ref{fig:projections}. An animation illustrating these two projections, and their effect on the observed column density distributions is available in the online version of the article, or at \href{https://youtu.be/3qbYvreT2AQ}{https://youtu.be/3qbYvreT2AQ}, and will greatly aid in the interpretation of this section. 

Each $\rm H_2$ column density projection is calculated using the \texttt{FITSOffAxisProjection} functionality from the \texttt{yt} package \citep{Turk_2011}. The \texttt{FITSOffAxisProjection} function integrates the $\rm H_2$ number density $\rm n_{H_2}$ along a line of sight $\hat{l}$, where the units of the projected field are equal to the units of the unprojected field times the appropriate length unit. The \texttt{OffAxis} component indicates that our line of sight $\hat{l}$ is an oblique angle and not perpendicular to any face of the cube. We assume the projection is centered on the central coordinate of each 500 pc$^3$ grid, with a resolution equal to the original cell size (1 pc), and a north vector perpendicular to the disk midplane, pointing towards positive z-values. The line of sight $\hat{l}$ of the observer is dependent on the two projection schemes and varies from filament to filament. In all cases, the observer is placed in the disc of the galaxy (z=0 pc), so we are only changing the orientation of the observer in the x-y plane (``top down" view) shown in Figure \ref{fig:projections}. 
 
The first projection is intended to produce a sample of filament properties largely free from projection effects (which can shorten filament lengths, etc.). Consider the box shown in the top panel of Figure \ref{fig:projections}. We start by projecting the data such that the vector pointing towards the observer is perpendicular to the front plane of the box (parallel to the vertical axis shown in Figure \ref{fig:projections}), so that the observer is facing towards the Galactic center. Then, following the procedure outlined in \S \ref{subsec:identification}, we identify filaments first in this projection. In order to estimate the properties of the filaments without projection effects, we then measure the 3D orientation of the filaments identified in 2D, by determining the angle between the long axis of the filament and the vector perpendicular to the front plane of the cube from a top-down perspective. The spine computed for one filament in the sample using this procedure is shown in blue inside the box in the top panel of Figure \ref{fig:projections}. After identifying the spine, we then project the cube again, so that the vector pointing towards the observer is perpendicular to the long-axis of the filament. The position of the observer in this scenario is shown via the spaceship in the top panel of Figure \ref{fig:projections}. We refer to this projection as the ``perpendicular" projection. 

The second projection is intended to simulate the properties we would observe given our position in the Galaxy. We do so with the ``sun-like" projection, determined by calculating the angle between the center of each box (shown in the bottom panel of Figure \ref{fig:pdfs}) and an assumed sun position in the Galaxy. The sun position is chosen in order to roughly replicate the distance and orientation of the near Scutum-Centaurus arm with respect to the Sun in the Milky Way Galaxy. The Scutum-Centaurus arm is the most prominent arm towards the inner Galaxy, at a distance of roughly 3-4 kpc. Over half of the sample of observed large-scale Galactic filaments lies closest to this arm \citep{Zucker_2018a}.  The most prominent spiral arm in the highly-resolved section of the \citet{Smith_2014} simulations lies at a y position of $\approx 4.0-4.5$ kpc, so a position of (11 kpc, 1 kpc) roughly mimics mimics this configuration. Our assumed sun position, (x,y) = (11 kpc, 1 kpc), is marked with a sun symbol in Figure \ref{fig:gridsearch}. As in the top panel of Figure \ref{fig:projections}, the orientation of the observer in the ``sun-like" projection is marked via a spaceship, with the tail of the spaceship oriented towards our assumed sun position. 

While in both projections we simulate the correct orientation of the observer with respect to the filament, it would be computationally prohibitive to account for column density unassociated with the filament along the line of sight. Thus, we only consider the column density locally around the filament, lying inside the same 500 pc$^3$ grid. This means that the synthetic filaments will have a much lower likelihood of confusion due to line-of-sight effects compared to observed filaments.

\subsection{Identifying Filaments} \label{subsec:identification}
A number of approaches to feature identification have been adopted in the literature, the most common being a dendrogram analysis \citep{Rosolowsky_2008, Colombo_2015} for molecular cloud identification, or the DisPerSE (DIScrete PERsistent Structures Extractor) algorithm \citep{Sousbie_2011} for identification of filamentary structure. However, neither approach is optimized for identifying elongated structures on large scales, and the dendrogram technique in particular has only been explicitly developed for use with quasi-spherical molecular clouds \citep{Rosolowsky_2008}. \citet{Duarte_Cabral_2016, Duarte_Cabral_2017} employ the SCIMES algorithm \citep{Colombo_2015}, which identifies molecular gas structures in dendrograms using the spectral cluster paradigm, which groups together discrete dendrogrammed regions with similar emission properties. Using SCIMES, \citet{Duarte_Cabral_2016, Duarte_Cabral_2017} identify only a few clouds with aspect ratios as high as $\approx 10:1$, which is even higher than the maximum aspect ratio identified for molecular clouds in observations \citep{Rice_2016} using only dendrograms with no spectral clustering. \citet{Li_2016} applies the DisPerSE algorithm to the ATLASGAL plane survey and identifies only a few filaments with lengths greater than 15 pc, with known large-scale Galactic filaments from other samples \citep{Wang_2015, Zucker_2015} being broken into multiple structures \citep[see also][]{Mattern_2018b}. 

Thus, in the past, the most common method for identifying large-scale Galactic filaments is to use simple column density or integrated intensity thresholds \citep{Zucker_2018a, Zucker_2015, Wang_2015, Abreu_Vicente_2016, Ragan_2014, Zhang_2019, Du_2017}. In almost all cases, initial filament selection is done by-eye in either dust extinction \citep[e.g. GLIMPSE \& MIPSGAL;][]{Churchwell_2009, Carey_2009} or dust emission \citep[e.g. Herschel;][]{Molinari_2016} before confirming velocity contiguity using either low ($\rm ^{13}CO$) or high ($\rm N_2H^+, NH_3$) density spectral-line tracers. This is also consistent with the by-eye identification of one of the largest molecular filaments in the Milky Way -- the 500 pc long ``Wisp" presented in \citet{Li_2013}. \citet{Li_2013} uses \texttt{Spitzer} 24 $\micron$ emission to track star formation in two molecular clouds, and combines it with $\rm ^{13}CO$ data from the GRS survey \citep{Jackson_2006}, to confirm that the double-cloud system actually belongs to an elongated molecular filament. 

\citet{Zucker_2018a} defines filament boundaries by setting a column density threshold $1-2 \sigma$ above the mean background column density of each Herschel Hi-GAL image \citep{Molinari_2016}, where the ``background" column density is defined using a low-emission region near the filament, following \citet{Juvela_2012}. Ideally, we would follow the same procedure as \citet{Zucker_2018a} to define boundaries. However, the dynamic range of the Hi-GAL column density maps spans a few orders of magnitude, while the dynamic range of the \citet{Smith_2014} AREPO simulations spans around twelve orders of magnitude. This is partly due to the very small scale-height of the disk, which results in steep dropoffs in column density beyond 20 pc from the midplane. The significantly larger dynamic range of the AREPO simulations makes the current observational approach for identifying filament boundaries challenging. While we do produce column densities for the filaments consistent with observations (see \S \ref{sec:results}), the role of feedback in setting the scale height of the disk cannot be underestimated. 

\begin{figure}[h!]
\begin{center}
\includegraphics[width=1.\columnwidth]{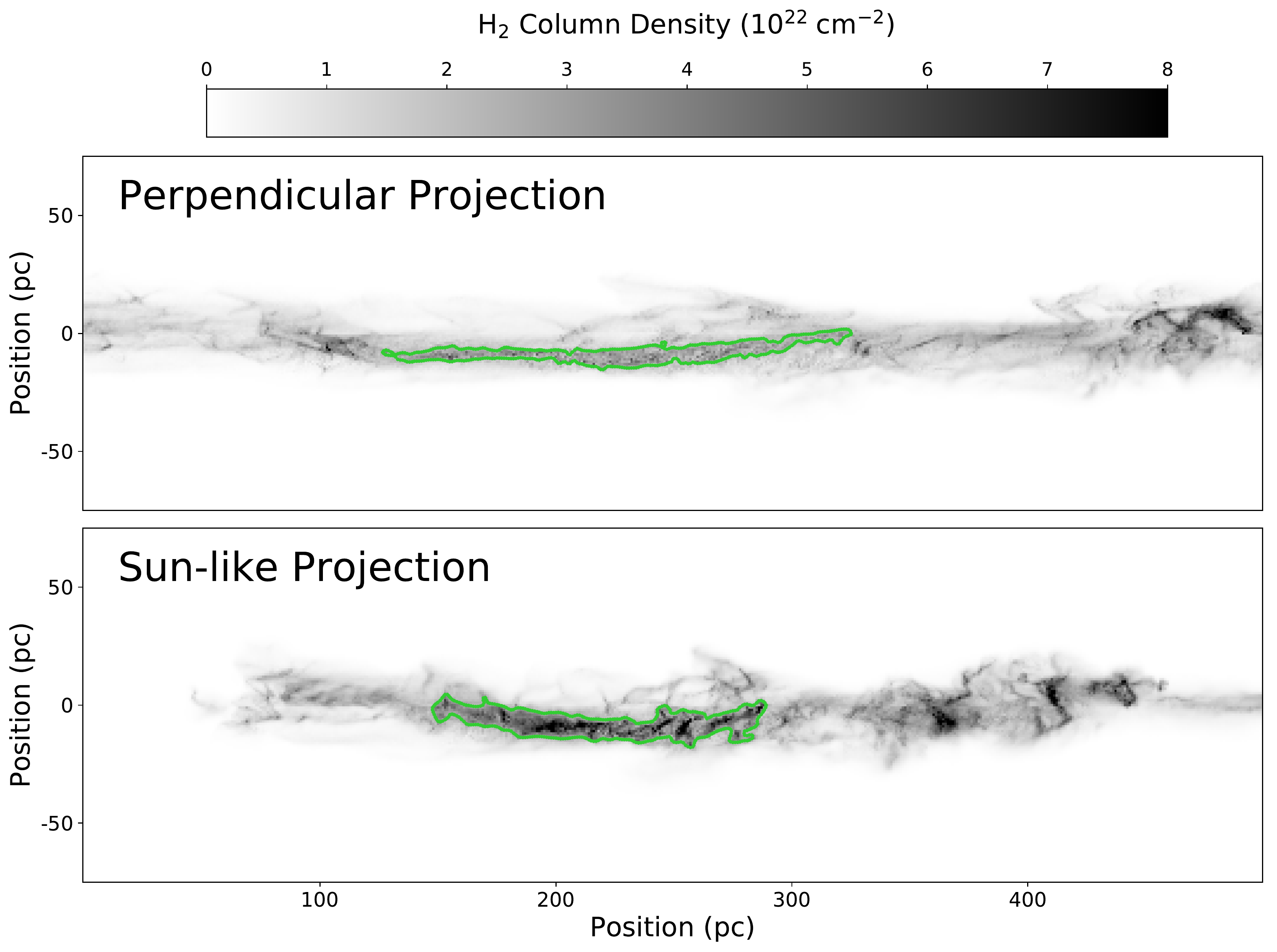}
\caption{ \label{fig:coldensity_projections} $\rm H_2$ column density map for the  ``perpendicular" projection (top) and the ``sun-like" projection (bottom) for ``Fil\_x916\_y400" (see Figure \ref{fig:projections}). The boundary defined in each projection is outlined in green. In the ``perpendicular" projection, the long axis of the filament is perpendicular to the line of sight of the observer, and should result in longer lengths and lower column densities. The ``sun-like" projection changes from filament to filament, and is dependent on how the filament is oriented in the synthetic galaxy with respect to our assumed sun position (x,y)=(11 kpc, 1 kpc); see Figure \ref{fig:projections}.}
\end{center}
\end{figure}

Thus, we adopt an alternative, very simple, but consistent approach to identifying the highest contrast filamentary features in each image. Specifically, we adopt the 99th percentile of column density over each image (the ``sun-like" and ``perpendicular" projections; see \S \ref{subsec:projections}) as the threshold for structure identification. We start by selecting all filaments (aspect ratio $\gtrsim$ 5)  first in the ``perpendicular" projection, where they will appear the longest to an observer. We then select the same filament in the ``sun-like" projection, in order to gauge the impact of projection effects on our analysis, and to simulate the range of physical properties we would observe given the sun's position in the galaxy. The boundaries of the same filament (``Fil\_x916\_y400") identified in both $\rm H_2$ column density projections are shown in green in Figure \ref{fig:coldensity_projections}. 
In practice, due to the absence of feedback and local self-gravity, the highest column density structures in these simulations are essentially entirely confined to large filaments, and line-of-sight confusion plays a negligible role, since we only account for structures inside each 500 pc$^{3}$ grid when performing the column density projections. Nevertheless, in both cases, we confirm velocity contiguity from an observer's perspective by taking a custom \textit{position-velocity} diagram over the filament's boundaries \citep{Zucker_2018a} with the glue visualization software \citep{glueviz_2017}. This is done using custom $\rm C^{18}O$ \textit{position-position-velocity} cubes computed for each filament, as discussed in \S \ref{subsec:radmc}.  Following the procedure outlined in \citet{Zucker_2018a}, if part of the filament is confirmed kinematically to be an unassociated structure along the line of sight, we mask out the intervening column density structure before reapplying the contour. As discussed in \S \ref{subsec:extraction}, since some filaments are split between grids, once the filaments are identified and confirmed, we shift the center of each grid (\S \ref{subsec:projections}) so that is aligned with the central coordinate of each filament. We repeat the same process described here and in \S \ref{subsec:projections} for the new grids, which results in modest changes to filament properties for structures located near the boundaries of the grids. The regridded projections are those we use to derive all filament properties described in \S \ref{sec:results}. 

\subsection{Radiative Transfer Modeling} \label{subsec:radmc}
We apply the gas line radiative transfer functionality in the software package RADMC-3D \citep{Dullemond_2012} to produce synthetic spectral cubes of the filaments identified in \S \ref{subsec:identification}. We use the $\rm C^{18} O \; (1-0)$ transition because it is abundant towards observed filaments in the literature (Battersby et al. 2019, in prep), yet optically thinner than the $\rm ^{12}CO$ and $\rm ^{13}CO$ lines. For computationaly expediency, we downsample the 1 pc$^3$ grids by a factor of two, and perform the radiative transfer modeling on grids with a spatial resolution of 2 pc$^{3}$. Otherwise, we adopt the same ``perpendicular" and ``sun-like" projections as used for the $\rm H_2$ column density maps.

RADMC-3D requires estimates of the dust temperature, dust density, gas temperature, gas velocity ($v_x, v_y, v_z$), $\rm C^{18}O$ abundance, collisional partner abundance, and the microturbulence in each cell.  We adopt molecular hydrogen as the collisional partner, as this is by far the most abundant molecule in molecular clouds. All these parameters except for microturbulence are extracted or derived from the output of the AREPO simulation. The dust density is derived from the gas density assuming a gas-to-dust ratio of 100:1. The $\rm C^{18}O$ abundance is derived from the AREPO CO abundance assuming an isotopic ratio of $\frac{\rm ^{12}CO}{\rm C^{18}O} = 557$ \citep{Wilson_1999}. While the microturbulence input is optional, we include it to account for spatially unresolved turbulent widths, on the order of $\rm 1 \; km \; s^{-1}$, as determined in \citet{Heyer_2004}.
 
RADMC-3D allows the user to run in several different line modes, including both LTE and non-LTE treatments. We adopt the Large Velocity Gradient (LVG) or ``Sobolev" mode, which is a non-LTE mode, as we cannot assume LTE in the typical diffuse conditions of the interstellar medium. If the source of interest undergoes large macroscopic motions, as the filaments typically do, the LVG mode allows one to make the approximation that the emission at one end of the cloud is completely decoupled from emission at the other end. This approximation is frequently applied to gas line radiative transfer modeling of molecular clouds, as described further in \citet{Ossenkopf_1997}. We turn on doppler catching, which prevents artifacts in the spectra in regions with large velocity gradients, due to the doppler shift in adjoining voxels exceeding the intrinsic linewidth of the material. 

These cubes are used to confirm velocity contiguity in identified filaments (as described in \S \ref{subsec:identification}) as well as in the velocity gradient analysis (described in \S \ref{sec:results}). 

\section{Results} \label{sec:results}
\subsection{Detection Statistics} \label{subsec:detection}
\begin{figure}[h!]
\begin{center}
\includegraphics[width=1.\columnwidth]{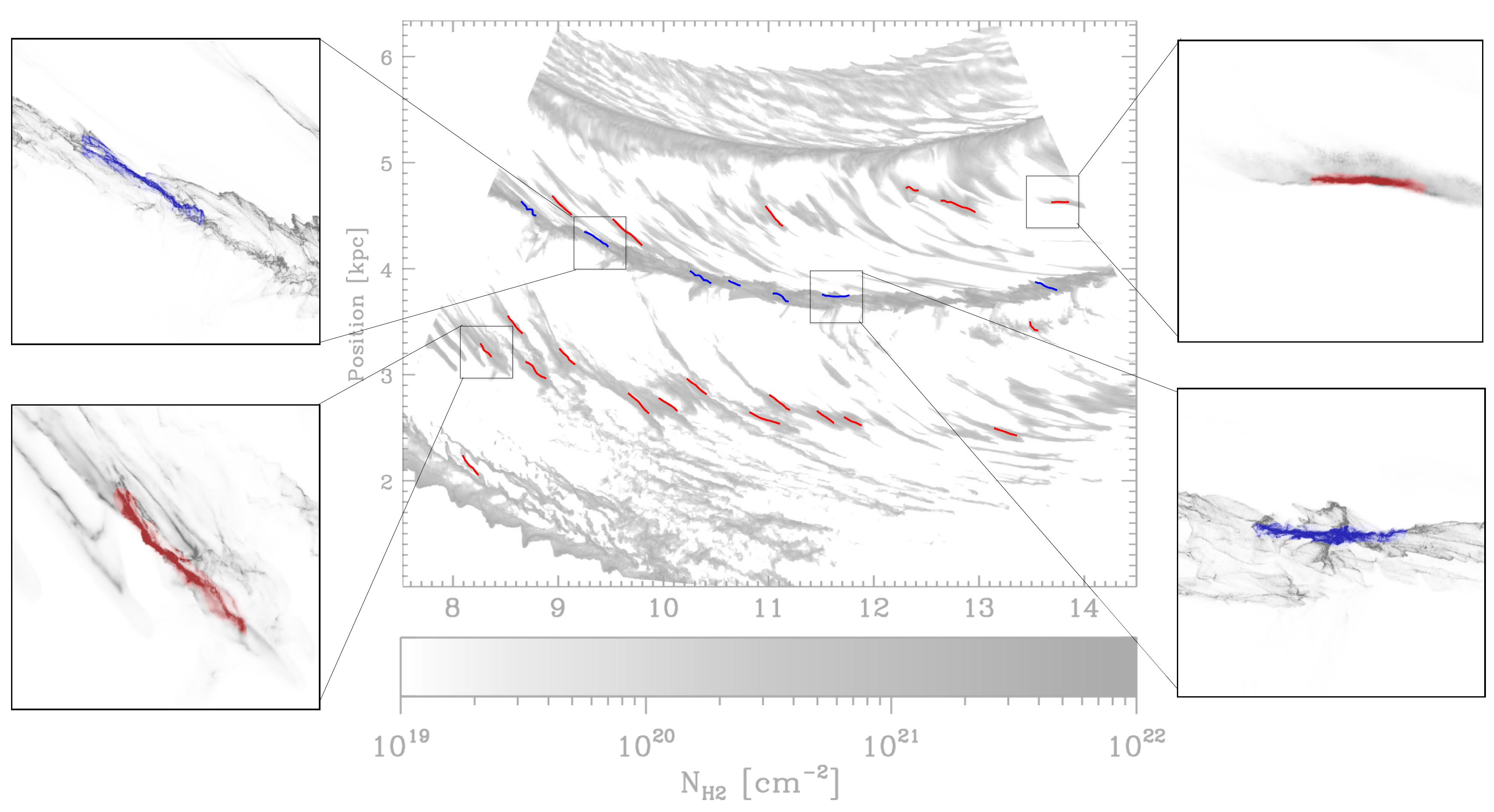}
\caption{ \label{fig:summary}Top down view of extracted filaments overlaid on the $\rm H_2$ column density map from \citet{Smith_2014}. ``Arm" filaments coinciding with the spiral-potential wells are shown in blue, while ``interarm" filaments lying in between the spiral-potential wells are shown in red. Zoomed-in regions showing the top-down $\rm H_2$ column density distribution for four different filaments are shown inside the pop-out boxes.}
\end{center}
\end{figure}

In Figure \ref{fig:summary} we show a top-down summary of the filaments identified following the procedure outlined in \S \ref{subsec:identification}. The spines of these filaments are delineated in blue if they are within 100 pc of the imposed spiral-potential wells. Otherwise, they are highlighted in red, indicating they lie between spiral-potential wells, in the interarm regions. 

In total we identify 27 filaments over the 31 grids we extract from the highly resolved section of the disk from \citet{Smith_2014}.  Recall that we only search for filaments over grids which contain appreciable amounts of CO emission, so an additional 41 grids (mostly in the interarm regions) are excluded based on this initial criterion. While all observed large-scale filaments are CO-bright, these 41 grids likely contain addition ``CO-dark" filaments which could potentially be observed using dust tracers. The total area of the zoom-in from the \citet{Smith_2014} simulation we targeted is $\rm \approx 20 \; kpc^{2}$, indicating that we should expect to identify $\approx 1-2$ filaments per square kiloparsec, assuming that the observer is in the disk of the galaxy. Extrapolating this to the entire Milky Way disk, and assuming a stellar disk with a radius of 15 kpc, we should expect to identify at maximum $\approx 1000$ filaments in the Milky Way. This estimate is in agreement with that made in \citet{Goodman_2014} (several hundreds to thousands), determined by comparing the mass in Nessie as traced by HNC to the total dense gas mass fraction in the Galaxy. 

It is important to emphasize that that these filaments are forming under ideal conditions, without stellar feedback or local self-gravity. Incorporating this additional physics should change the number of observable filaments. With the inclusion of the local self-gravity and feedback, these structures would quickly collapse and form high-mass stars. Feedback from the stars formed would presumably break filaments into discrete pieces on short timescales ($\approx$ a few million years), and also push them farther from their birthplaces in the gravitational midplane. In counting filaments, stellar feedback has the potential to both increase or decrease tallies. If a 300 pc long filament in the spiral arm is broken apart by feedback, it is possible that sections of this filament would still be identifiable as one or more distinct, smaller scale filaments ($\approx$ tens to a hundred parsecs in length). However, in many cases, the filament may likely also be completely destroyed, and no longer identifiable from an observer's perspective. This is likely also time and location dependent, and could depend on the filament's environment, the number of high-mass stars, and their distribution along the filament. 

Clumps of high-density gas, alongside nascent HII regions, are observable towards large-scale Galactic filaments in our own Milky Way \citep[see][]{Jackson_2010, Tackenberg_2013, Wang_2014}. The HII regions make these features harder to identify as a coherent filamentary structure as they evolve. Since the formation of these filaments is dominated by Galactic dynamics \citep[rather than local self-gravity, as we see for nearby filaments;][]{Hacar_2013, Arzoumanian_2011, Andre_2010, Hacar_2018} these large-scale filaments should constantly be forming and/or persisting in ``special" places in the galaxy (either in the potential wells of spiral arms, or in the interarm regions). Since we only analyze a single time snapshot in the current work, we should be able to better characterize this in future work, by tracking the evolution of these filaments over time as they cross the spiral potential wells \citep[as in][]{Duarte_Cabral_2017}. This is discussed in \S \ref{subsec:formation}. 

\subsection{Potential Formation Mechanisms} \label{subsec:formation}
In Figure \ref{fig:summary} we find two classes of filaments: one type (highlighted in blue) lies inside the spiral-potential wells, while the other type (highlighted in red) lies between spiral-potential wells. We identify seven filaments along spiral arms, and twenty in the interarm regions, though these statistics are skewed by the fact we include two interarm regions in the search area, and only a single spiral arm. Nevertheless, this is broadly consistent with the findings of \citet{Zucker_2018a} -- who find that while most filaments lie in the plane of the Milky Way, only about $30\%$ of large-scale filaments also lie at velocities which are kinematically consistent with known spiral arms models \citep[see][]{Reid_2016}.

These two classes of filaments --- ``arm" and ``interarm" --- may correspond to two unique formation mechanisms. The filaments in blue may form as gas enters the spiral-potential wells, and becomes shocked and compressed, leading to high-contrast yet transient filamentary features forming along the arms. The filaments in red may form due to differential rotation, as gas is stretched out into lower column density structures in the interarm regions. While most of the molecular hydrogen in the interarm regions is CO dark, the emission coincident with the large-scale filaments we identify are ``islands" of CO bright emission, as they represent the density peaks of the interarm gas \citep[c.f. Figure 5 in][]{Smith_2014}. Interestingly, while the interarm filaments may form outside the potential wells, they appear to form at the same Galactocentric radius, with most of the interarm filaments forming within $\approx 200$ pc of y=3 kpc in the context of Figure \ref{fig:summary}. When stitched together, these filaments could easily be mistaken for an ``arm-like" structure. In this sense, the interarm filaments are tracing a larger, potentially transient, galactic-scale structure that develops dynamically in the simulations and often resembles an ``arm". 

Ultimately, however, in order to constrain the formation mechanisms of the filaments, we must track the evolution of the gas over megayears, as opposed to a single time snapshot, as we do in this work. In future work, we plan to build upon the time evolution analysis presented in \citet{Duarte_Cabral_2017}, which tracks two GMFs over eleven megayears, as they undergo interarm passage and cross a spiral potential well. This in turn will shed light on whether the properties of the filaments are set by their formation or their current dynamical environment. This is an important distinction, given the slow formation of CO-dominated regions in molecular clouds \citep{Clark_2012}, and our criterion that the synthetic filaments be CO bright. 

\subsection{Physical Properties}
To determine the physical properties of the synthetic large-scale filaments (length, width, mass, linear mass, column density, position angle, galactic plane separation, and velocity gradient), we adopt the same methodology applied to observed large-scale filaments in \citet{Zucker_2018a}. 

In Table \ref{tab:sumtab}, we summarize the physical properties for the full sample, calculated over both the ``perpendicular" and ``sun-like" projections. We briefly describe the method used to calculate each property in \citet{Zucker_2018a} in subsections below, along with the dependence of each property on projection effects and galactic environment (e.g. whether the filament lies in an arm or an interarm region). A machine readable version of Table \ref{tab:sumtab} is available at the Harvard Dataverse (\href{https://doi.org/10.7910/DVN/SPX2LL}{doi:10.7910/DVN/SPX2LL})

Due to the simplified physics involved many of these properties should be considered upper or lower limits on the actual values we would observe if feedback and local self-gravity were included. In Table \ref{tab:caution}, we include details on which properties are known to be biased, and also highlight whether the values we calculate should be treated as an upper or lower limit, along with the physics (or lack thereof) that is driving the bias itself (e.g. no feedback, no local self-gravity). 

\begin{deluxetable}{@{\extracolsep{0pt}}cccccccccccccccccc}
\tablecaption{Synthetic Filament Properties \label{tab:sumtab}}
\renewcommand{\arraystretch}{1.}
\colnumbers
\tabletypesize{\scriptsize}
\tablehead
{
\rule{0pt}{4ex}    
\multirow{5}{*}{\textbf{ID}} & \multirow{5}{*}{\textbf{Environment}} & \multicolumn{2}{c}{Length} & \multicolumn{2}{c}{Width} & \multicolumn{2}{c}{$\rm N_{H_2}$} & \multicolumn{2}{c}{Mass} & \multicolumn{2}{c}{Line Mass} &  \multicolumn{2}{c}{$z$} & \multicolumn{2}{c}{$\theta$} &  \multicolumn{2}{c}{$\frac{dv}{dl}$}\\
& & \multicolumn{2}{c}{pc} & \multicolumn{2}{c}{pc} & \multicolumn{2}{c}{$\rm cm^{-2}$} & \multicolumn{2}{c}{$\rm M_\sun$} & \multicolumn{2}{c}{$\rm M_\sun \; pc^{-1}$} &  \multicolumn{2}{c}{pc} & \multicolumn{2}{c}{$\circ$} & \multicolumn{2}{c}{$\rm km\; s^{-1} \; pc^{-1}$} \\[4.5pt]
 \cline{3-4} \cline{5-6} \cline{7-8} \cline{9-10} \cline{11-12} \cline{13-14} \cline{15-16} \cline{17-18}  \\[-5pt] 
&  & Perp & Sun & Perp & Sun & Perp & Sun  & Perp & Sun  & Perp & Sun  & Perp & Sun  & Perp & Sun & Perp & Sun}
\startdata
x1000y265 & Interarm & 163 & 89 & 10 & 6 & 5.5e+21 & 6.8e+21 & 2.0e+05 & 1.0e+05 & 1.2e+03 & 1.2e+03 & 20 & 19 & 1 & 4 & 0.027 & 0.017 \\
x945y255 & Interarm & 188 & 87 & 11 & 20 & 6.0e+21 & 9.5e+21 & 3.1e+05 & 3.9e+05 & 1.6e+03 & 4.4e+03 & 4 & 4 & 1 & 8 & 0.022 & 0.049 \\
x887y447 & Interarm & 343 & 202 & 9 & 12 & 3.2e+21 & 4.3e+21 & 1.9e+05 & 2.8e+05 & 5.6e+02 & 1.4e+03 & 4 & 4 & 0 & 1 & 0.012 & 0.039 \\
x888y300 & Interarm & 213 & 128 & 11 & 22 & 5.6e+21 & 1.1e+22 & 3.5e+05 & 6.9e+05 & 1.6e+03 & 5.4e+03 & 1 & 3 & 0 & 5 & 0.014 & 0.009 \\
x848y284 & Interarm & 281 & 91 & 9 & 28 & 6.4e+21 & 1.2e+22 & 3.8e+05 & 8.2e+05 & 1.4e+03 & 8.9e+03 & 5 & 6 & 1 & 12 & 0.015 & 0.029 \\
x830y330 & Interarm & 271 & 75 & 10 & 34 & 3.3e+21 & 5.3e+21 & 2.1e+05 & 4.6e+05 & 7.7e+02 & 6.0e+03 & 4 & 5 & 1 & 15 & 0.012 & 0.063 \\
x800y200 & Interarm & 167 & 94 & 8 & 11 & 5.6e+21 & 8.7e+21 & 1.9e+05 & 2.5e+05 & 1.1e+03 & 2.6e+03 & 9 & 8 & 7 & 16 & 0.034 & 0.021 \\
x808y295 & Interarm & 241 & 124 & 9 & 18 & 6.9e+21 & 1.3e+22 & 3.7e+05 & 7.9e+05 & 1.5e+03 & 6.4e+03 & 9 & 5 & 3 & 9 & 0.016 & 0.004 \\
x1350y435 & Interarm & 247 & 242 & 11 & 11 & 1.8e+21 & 1.8e+21 & 1.1e+05 & 1.3e+05 & 4.5e+02 & 5.2e+02 & 3 & 3 & 2 & 2 & 0.017 & 0.005 \\
x1330y310 & Interarm & 150 & 141 & 15 & 16 & 2.4e+21 & 2.5e+21 & 1.3e+05 & 1.3e+05 & 8.6e+02 & 9.3e+02 & 24 & 24 & 1 & 0 & 0.007 & 0.005 \\
x950y420 & Interarm & 442 & 239 & 7 & 9 & 5.1e+21 & 8.4e+21 & 4.3e+05 & 5.1e+05 & 9.7e+02 & 2.1e+03 & 1 & 1 & 0 & 0 & 0.014 & 0.037 \\
x1312y220 & Interarm & 182 & 181 & 14 & 16 & 3.8e+21 & 5.1e+21 & 2.1e+05 & 3.2e+05 & 1.2e+03 & 1.8e+03 & 5 & 4 & 0 & 2 & 0.011 & 0.004 \\
x987y235 & Interarm & 216 & 137 & 12 & 12 & 9.5e+21 & 1.1e+22 & 5.6e+05 & 3.9e+05 & 2.6e+03 & 2.8e+03 & 1 & 2 & 0 & 2 & 0.018 & 0.055 \\
x1213y448 & Interarm & 168 & 171 & 15 & 15 & 3.4e+21 & 3.3e+21 & 2.2e+05 & 2.2e+05 & 1.3e+03 & 1.3e+03 & 15 & 15 & 2 & 2 & 0.024 & 0.023 \\
$\rm x1140y225_1$ & Interarm & 132 & 128 & 5 & 5 & 7.8e+21 & 7.6e+21 & 1.2e+05 & 1.2e+05 & 9.3e+02 & 9.7e+02 & 4 & 4 & 1 & 1 & 0.009 & 0.012 \\
$\rm x1140y225_2$ & Interarm & 163 & 160 & 11 & 11 & 9.0e+21 & 8.6e+21 & 3.7e+05 & 3.6e+05 & 2.3e+03 & 2.3e+03 & 1 & 1 & 2 & 2 & 0.016 & 0.018 \\
x1080y425 & Interarm & 297 & 267 & 6 & 7 & 3.1e+21 & 3.3e+21 & 1.4e+05 & 1.6e+05 & 4.7e+02 & 6.1e+02 & 2 & 3 & 1 & 1 & 0.017 & 0.034 \\
x1087y260 & Interarm & 236 & 226 & 10 & 12 & 4.7e+21 & 5.2e+21 & 2.8e+05 & 3.1e+05 & 1.2e+03 & 1.4e+03 & 20 & 20 & 3 & 3 & 0.017 & 0.025 \\
x1066y218 & Interarm & 205 & 179 & 9 & 13 & 3.1e+21 & 3.5e+21 & 1.4e+05 & 1.8e+05 & 7.0e+02 & 1.0e+03 & 10 & 10 & 0 & 0 & 0.014 & 0.018 \\
x1258y432 & Interarm & 336 & 322 & 7 & 7 & 7.6e+21 & 8.0e+21 & 5.2e+05 & 4.9e+05 & 1.5e+03 & 1.5e+03 & 3 & 3 & 2 & 2 & 0.024 & 0.021 \\
x1138y347 & Arm & 272 & 297 & 8 & 7 & 3.4e+22 & 3.3e+22 & 2.2e+06 & 2.3e+06 & 7.9e+03 & 7.6e+03 & 5 & 5 & 0 & 0 & 0.024 & 0.024 \\
x843y434 & Arm & 150 & 80 & 11 & 12 & 2.3e+22 & 5.0e+22 & 9.0e+05 & 1.2e+06 & 6.0e+03 & 1.5e+04 & 2 & 1 & 0 & 1 & 0.034 & 0.055 \\
x1089y347 & Arm & 135 & 100 & 10 & 8 & 3.0e+22 & 3.0e+22 & 9.9e+05 & 6.1e+05 & 7.3e+03 & 6.1e+03 & 5 & 7 & 7 & 9 & 0.020 & 0.044 \\
x1049y358 & Arm & 134 & 94 & 6 & 4 & 1.5e+22 & 1.7e+22 & 2.9e+05 & 1.9e+05 & 2.1e+03 & 2.0e+03 & 5 & 5 & 0 & 0 & 0.003 & 0.035 \\
x916y400 & Arm & 222 & 163 & 4 & 8 & 2.8e+22 & 4.1e+22 & 6.7e+05 & 1.3e+06 & 3.0e+03 & 8.2e+03 & 7 & 7 & 2 & 1 & 0.004 & 0.036 \\
x1011y365 & Arm & 221 & 152 & 7 & 9 & 3.7e+22 & 4.9e+22 & 1.6e+06 & 1.7e+06 & 7.3e+03 & 1.1e+04 & 12 & 13 & 0 & 0 & 0.018 & 0.031 \\
x1338y357 & Arm & 315 & 241 & 6 & 9 & 4.2e+22 & 5.4e+22 & 2.3e+06 & 2.7e+06 & 7.3e+03 & 1.1e+04 & 2 & 2 & 0 & 0 & 0.021 & 0.011 
\enddata
\tablecomments{A summary of the physical properties we calculate for the full sample shown in Figure \ref{fig:summary}. For each property, we show the value determined assuming two different observer positions---the ``perpendicular" (``perp") and ``sun-like" (``sun") projections; see \S \ref{subsec:projections} or Figure \ref{fig:projections}. In (1) we list the ID for the filament, defined using the lower left hand corner position of the AREPO grid it is identified on in the right panel of Figure \ref{fig:gridsearch} (e.g. x1011y365 is equivalent to x=10.11 kpc, y=3.65 kpc in the context of Figure \ref{fig:gridsearch}). In (2) we list whether the filament is located in an arm or interarm region. In (3) and (4) we list the length of the filament for the perpendicular and sun-like projections. In (5) and (6) we list the width of the filament for the perpendicular and sun-like projections. In (7) and (8) we list the median $\rm H_2$ column density inside each filament's mask for both the perpendicular and sun-like projections. In (9) and (10) we list the mass of the filament for both the perpendicular and sun-like projections. In (11) and (12) we list the linear mass of the filament for both the perpendicular and sun-like projections. In (13) and (14) we list the 2D projected separation between the filament and the midplane, for both the perpendicular and sun-like projections. In (15) and (16) we list the 2D projected orientation between the long axis of the filament and the midplane, for both the perpendicular and sun-like projections. Finally, in (17) and (18) we list the velocity gradient along the filament, for both the perpendicular and sun-like projections. A machine readable version of this table is available on the Harvard Dataverse (\href{https://doi.org/10.7910/DVN/SPX2LL}{doi:/10.7910/DVN/SPX2LL})}
\end{deluxetable}

\begin{deluxetable*}{ccc}
\tablecaption{Potentially Biased Physical Properties\label{tab:caution}}
\tablewidth{0pt}
\tablehead{
\colhead{Property} & \colhead{Upper Limit/Lower Limit?} & \colhead{Reason?}}
\startdata
Length & Upper Limit & No Feedback \\
Width & Upper Limit & No Local Self-Gravity \\
Mass & Upper Limit & No Feedback \\
Plane Separation & Lower Limit & No Feedback \\
Position Angle & Lower Limit & No Feedback
\enddata
\end{deluxetable*}

\subsubsection{Lengths \& Widths} \label{subsubsec:lengths}
To compute the lengths, we utilize the boundaries defined for the filaments in \S \ref{subsec:identification}. Specifically, the area inside each boundary constitutes a mask, which is skeletonized using a medial axis transform via the \texttt{FilFinder} package \citep{Koch_2015}. The resulting skeleton or ``spine" represents a one pixel wide geometric representation of the mask topology, and the length of this spine is the length we adopt for the filament. 

We compute the widths of the filaments using the \texttt{RadFil} package \citep{Zucker_2018b}. The \texttt{RadFil} package computes filament widths by taking perpendicular cuts across the spine of the filament. We compute the geometric width of each filament, by determining where the cut touches the edge of the mask on either side of the spine. Each cut samples the ``local" width of the filament determined by the column density threshold. We then take the median of all these cuts to compute the contour-based width of the filament.\footnote{Note this is a different procedure than employed in \citet{Zucker_2018a}. In that work we fit a Gaussian and Plummer function to the radial column density profile. However, this is challenging to replicate for the synthetic filament population for two reasons. First, the filaments are more flocculent than those in the Milky Way, and second, the large dynamic range (twelve orders of magnitude) and small scale height of the disk in the AREPO simulation make it difficult to fit and subtract the background column density}

\begin{figure}[h!]
\begin{center}
\includegraphics[width=1. \columnwidth]{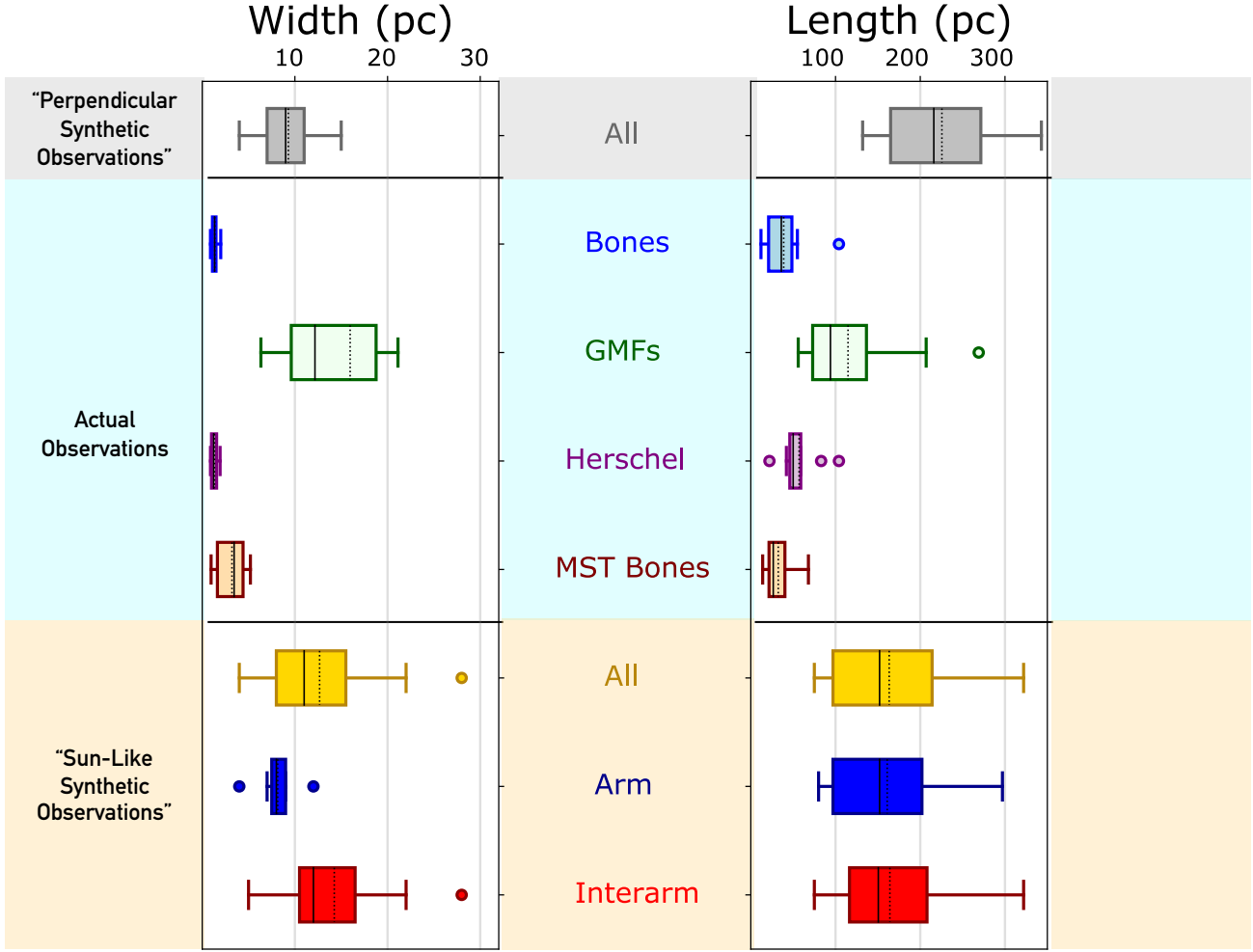}
\caption{ \label{fig:lengths} \textbf{Left}: Box and Whisker plots of the filament width distribution, over three parameter spaces. In the top left panel, we show the distribution of widths for the full sample from the ``perpendicular" projection (gray). In the middle left panel, we show the observed distribution of filament widths.  In the bottom left panel, we show the distribution of widths for the full sample from the ``sun-like" projection (yellow), as well as for the arm (blue) and interarm (red) sub-samples in the same sun-like projection. \textbf{Right}: Same as in the left panel, but for the distribution of lengths. In both panels the solid line marks the median of the distribution, while the dashed line indicates the mean.}
\end{center}
\end{figure}

In Figure \ref{fig:lengths} we compare (via box-and-whisker plots) how the distribution of lengths varies due to projection effects and galactic environment. We find that on average, the filament lengths tend to be foreshortened $1.4\times$ due to projection effects, with the median length in the ``perpendicular" projection of 216 pc, and in the ``sun" projection of 152 pc. We find no difference in the observable length of the filaments due galactic environment, with a typical projected filament length in the arm region of 152 pc, and in the interarm region of 150 pc. In all cases, the synthetic filaments are typically $1.5 - 10 \times$ longer than what we observe in our own Galaxy \citep{Zucker_2018a}. The lowest density class of observed large-scale filament (the ``GMFs"), with lengths around 100 pc, are most consistent with the lengths of the synthetic filaments. The densest classes of large-scale filaments (the ``Bones" and ``Herschel" filaments), with lengths of around 50 pc, are typically shorter by a factor of three.\footnote{The lengths of the ``Bone-like" filaments \citep{Zucker_2018a} are typically lower limits, given our methodology of requiring semi-continuous closed contours despite the presence of HII regions across the filaments. For example, \citet{Zucker_2018a} find a length of Nessie of 104 pc, whereas \citet{Goodman_2014} find the length of Nessie to be between 160-430 pc, depending on how generously one connects the different filamentary extinction features} The distribution of synthetic filament lengths in all cases should be considered an upper limit, as the incorporation of feedback (from HII regions or supernovae) should break the filaments into pieces, making it harder to identify them as coherent structures from an observer's perspective. 

Also in Figure \ref{fig:lengths}, we show a box-and-whisker diagram for the distribution of filament widths. The widths of the filaments are less dependent on viewing angle, with the typical width in the ``sun-like" projection of 11 pc, only a few parsecs larger than the ``perpendicular" projection (9 pc). The difference in width based on environment is modestly more pronounced, with filaments forming in the arm having moderately smaller widths (8 pc) compared to interarm filaments (12 pc). The widths we observe for the GMFs in our Galaxy ($\approx 13$ pc) are most consistent with the synthetic filaments, particularly for interarm filaments seen from the ``sun-like" projection. These simulations are unable to reproduce the widths of the denser filament catalogs -- including Nessie -- which have typical widths of $\approx 1-2$ pc. The simulated widths should be taken as upper limits, due to the lack of local self-gravity in the simulations. The inclusion of local self-gravity should cause radial collapse along the filaments, narrowing not only each filament's column density profile, but also the ``mask-based" width we calculate based on an assumed column density threshold, as the contrast between filament and background should be higher in these cases. 

\subsubsection{$H_2$ Column Densities, Masses, \& Linear Masses}
Column densities, masses, and linear masses are calculated using the $\rm H_2$ column density projections described in \S \ref{subsec:projections}. We compute the median $\rm H_2$ column density for each filament by taking the median column density value inside each filament mask. To compute the masses following \citet{Zucker_2018a}, we take the integral of the $\rm H_2$ column densities across the filament mask ($\rm M_{tot}=\mu_{H_2} m_H \int N_{H_2} dA$), where $\rm \mu_{H_2}=2.8$ \citep{Kauffmann_2008} and the integral is approximated by taking the sum over the $\rm H_2$ column density of each pixel times its physical area (1 pc$^2$). The linear masses are obtained by dividing these masses by the lengths we compute in the previous section.

\begin{figure}[h!]
\begin{center}
\includegraphics[width=0.8 \columnwidth]{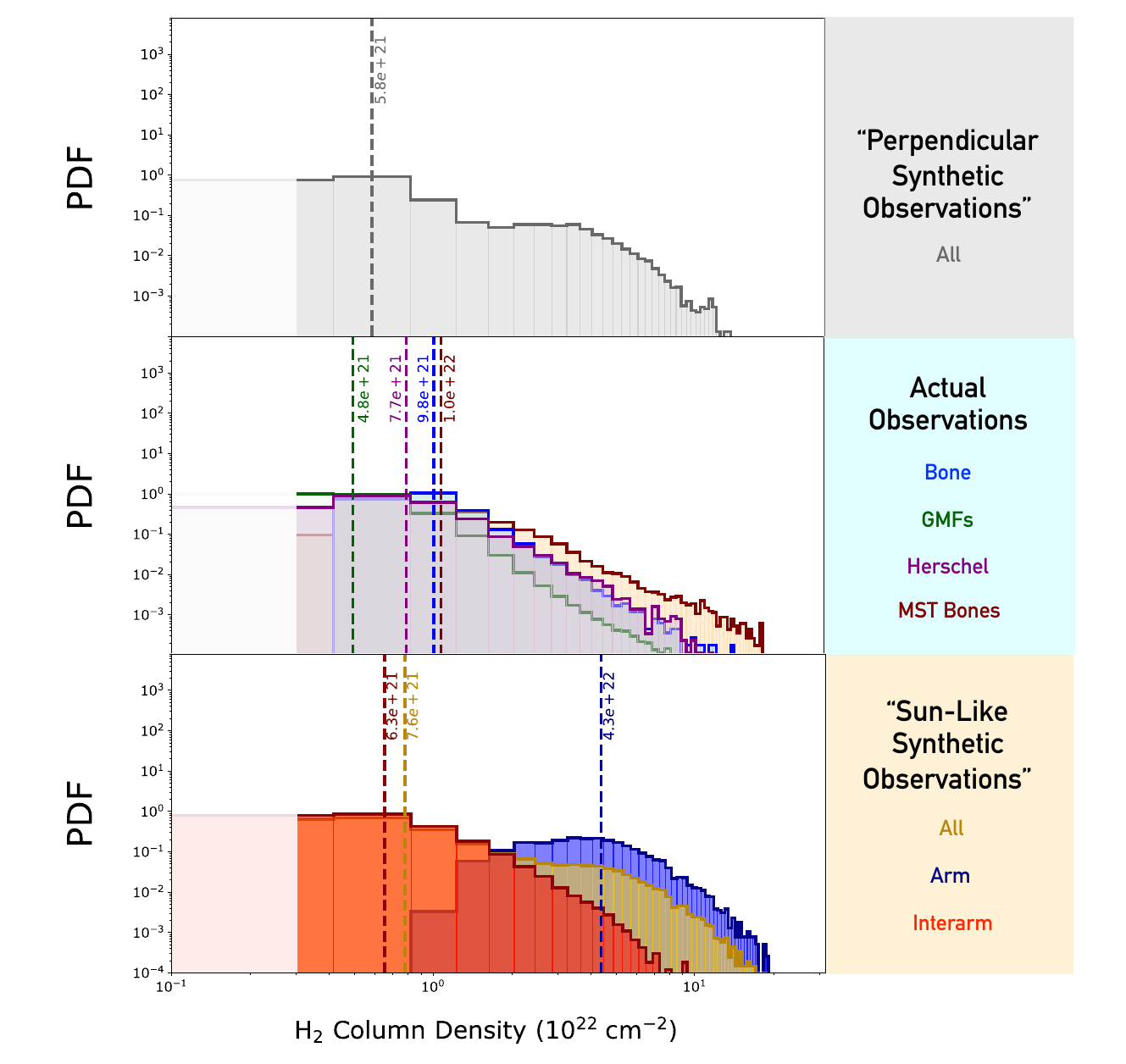}
\caption{ \label{fig:pdfs} $\rm H_2$ column density PDFs showing the column density distribution for each
filament class as a whole. Each PDF is normalized such that its total area is equal to one. The ``perpendicular observer" (top panel) and ``sun-like observer" (bottom panel) classes show the distribution of  the full sample as seen from the perpendicular and sun-like projections. The ``arm" and ``interarm" classes (bottom panel) show the distribution of column densities for filaments in the arm and interarm regions, as seen from the ``sun-like" projection. The middle panel shows the observed distributions. The median column density distribution of each class is labeled and marked with the vertical dashed lines. }
\end{center}
\end{figure}

Column density PDFs are summarized in Figure \ref{fig:pdfs}. For each distribution, we stack all the pixels inside each filament's mask, to obtain class-by-class statistics. 
As shown in Figure \ref{fig:pdfs}, we find a modest increase in the $\rm H_2$ column density due to projection effects, with the median column density for the ``sun-like" projection being $30\%$ higher than in the perpendicular projection ($\rm 7.6\times10^{21} \; cm^{-2}$ vs. $\rm 5.8\times10^{21} \; cm^{-2}$). Given that we do not account for intervening column density along the line of sight, this should be treated as an ideal case, which is only dependent on the viewing angle. In contrast, the difference in column density between the arm and interarm filaments is quite pronounced, with the median column density almost a factor of seven higher in the arm ($\rm 4.3\times10^{22} \; cm^{-2}$) vs. the interarm regions ($\rm 6.3\times10^{21} \; cm^{-2}$). As a whole, the column densities of the synthetic filaments agree well with those measured in our own Galaxy. The typical column densities of observed filaments varies from $\rm 4.8\times10^{21} \; cm^{-2}$ for the lowest column density class (the ``GMFs") to $\rm 1\times10^{22} \; cm^{-2}$ for the highest column density class (the ``Bones" and ``MST Bones"). Thus, like the widths, the GMFs tend to be more consistent with the ``interarm" filament column densities, while the denser filament categories tend to be more consistent with -- but not as high as -- the arm filament column densities. The good \textit{average} agreement between observations and simulations, despite the fact that we only consider the column density locally around each filament, indicates that the observed large-scale filaments may dominate the total column density along the line of sight. The column densities of the observed filaments are beam-diluted given the resolution \citep[see discussion in \S 3.2 of][]{Zucker_2018a} so any build-up in intervening column density along the line of sight in our own Galaxy could be compensated for by the beam dilution inherent in the column density measurements we report in \citet{Zucker_2018a}. While the average column densities are similar, the \textit{shapes} of the column density PDFs show significantly less agreement -- the simulations lack the characteristic ``power law tail" \citep{Ballesteros_Paredes_2011, Burkhart_2017}, likely due to the absence of local self-gravity. 

\begin{figure}[h!]
\begin{center}
\includegraphics[width=1. \columnwidth]{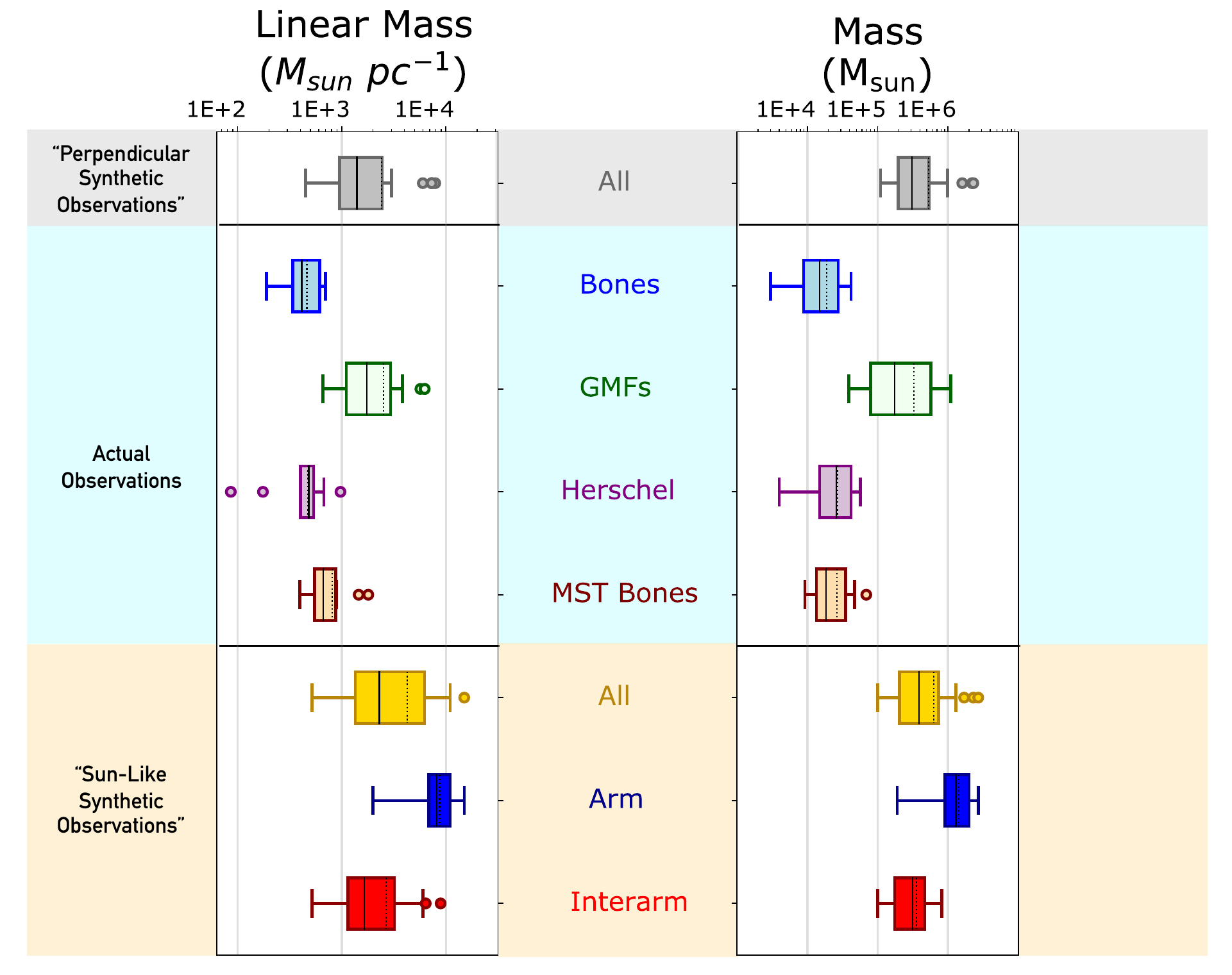}
\caption{ \label{fig:mass} Same as Figure \ref{fig:lengths}, but for the distribution of linear masses (left) and masses (right).}
\end{center}
\end{figure}

In Figure \ref{fig:mass} we show the comparisons for mass and linear mass, over the same parameter space explored in Figure \ref{fig:lengths}. We find that the mass is only modestly influenced by projection effects, with the ``sun-like" projection having a typical mass of $\rm 3.9\times10^{5} \; M_\sun$, vs. $\rm 3.1\times10^{5} \; M_\sun$ for the ``perpendicular" projection. Due to their increased $\rm H_2$ number densities, and thus increased $\rm H_2$ column densities, the masses of the arm filaments ($\rm 1.3 \times 10^{6} \; M_\sun$) are $4 \times$ higher than the interarm filaments ($\rm 3.2 \times 10^{5} \; M_\sun$). The masses of the synthetic filaments are typically one to two orders of magnitude higher than what we observe in our Galaxy, with most large-scale Galactic filaments masses lying just above $\rm 10^{4} \; M_\sun$. The exception is the ``GMF" class, which has a typically mass of $\rm 10^{5} \; M_\sun$, in good agreement with the synthetic interarm filament distribution we calculate. Also in Figure \ref{fig:lengths}, we find that, unsurprisingly, the influence of projection effects on the linear mass follows the same trend as for length, with the ``sun-like" projection ($\rm 2.3\times10^{3} \; M_\sun \; pc^{-1}$), having a linear mass on average $1.6\times$ higher than for the ``perpendicular" projection ($\rm 1.4\times10^{3} \; M_\sun \; pc^{-1}$). Like mass and column density, the linear mass is clearly affected by galactic environment, with filaments in the arm having linear masses $5\times$ higher than interarm filaments ($\rm 8.2\times10^{3} \; M_\sun \; pc^{-1}$ and $\rm 1.6\times10^{3} \; M_\sun \; pc^{-1}$, respectively). As for mass, the linear masses of the GMFs observed in our Galaxy, typically lying around $\rm 1.5 \times 10^{3} \; M_\sun \; pc^{-1}$, show good agreement with the synthetic ``interarm" filaments. Non-GMF observed large-scale filaments, including filaments like Nessie, have lower linear masses, on the order of $\rm \approx 500 \; M_\sun \; pc^{-1}$.

Like the length and width, we expect the incorporation of local self-gravity and feedback to affect the masses and linear masses we infer. Similar to the lengths, the inclusion of feedback will break apart and potentially destroy entirely sections of filaments, which could reduce their mass significantly. If and how much the linear mass decreases depends on whether feedback affects both the length and the mass the same, or whether one property is preferentially affected. Concurrently, the inclusion of local self-gravity will likely narrow the filaments (potentially up to a factor of ten, if they are to be consistent with observations) and also increase the median column densities we infer. Whether that would significantly change the linear masses and column densities will depend on how the decreased surface area of the filaments balances the anticipated increase in column density.

\subsubsection{Galactic Plane Separation \& Position Angle}
We calculate the the galactic plane separation and position angle using the location of the gravitational midplane in the \citet{Smith_2014} simulations (the z=0 pc axis).  The galactic plane separation is the 2D projected separation between the filament and the disk midplane. We compute it following the methodology outlined in \citet{Zucker_2018a}; that is we find the minimum absolute separation between every point on the filament's spine and the closest point in the midplane. The median of these minima is the plane separation we adopt for the filament. To compute the position angle, we fit a line to each filament's mask, assuming every pixel in the mask is a scatter point. The filament position angle can be parameterized as $\theta_{fil} = \arctan{(m_{fil})}$ and the plane position angle can be parameterized as $\theta_{plane} = \arctan{(m_{plane})}$, where $m_{fil}$ is the slope of the line fitted to the filament's mask, and $m_{plane}$ is the slope of the plane (assumed to be zero). The position angle we report is the absolute difference between the filament position angle and the plane position angle. 

\begin{figure}[h!]
\begin{center}
\includegraphics[width=1. \columnwidth]{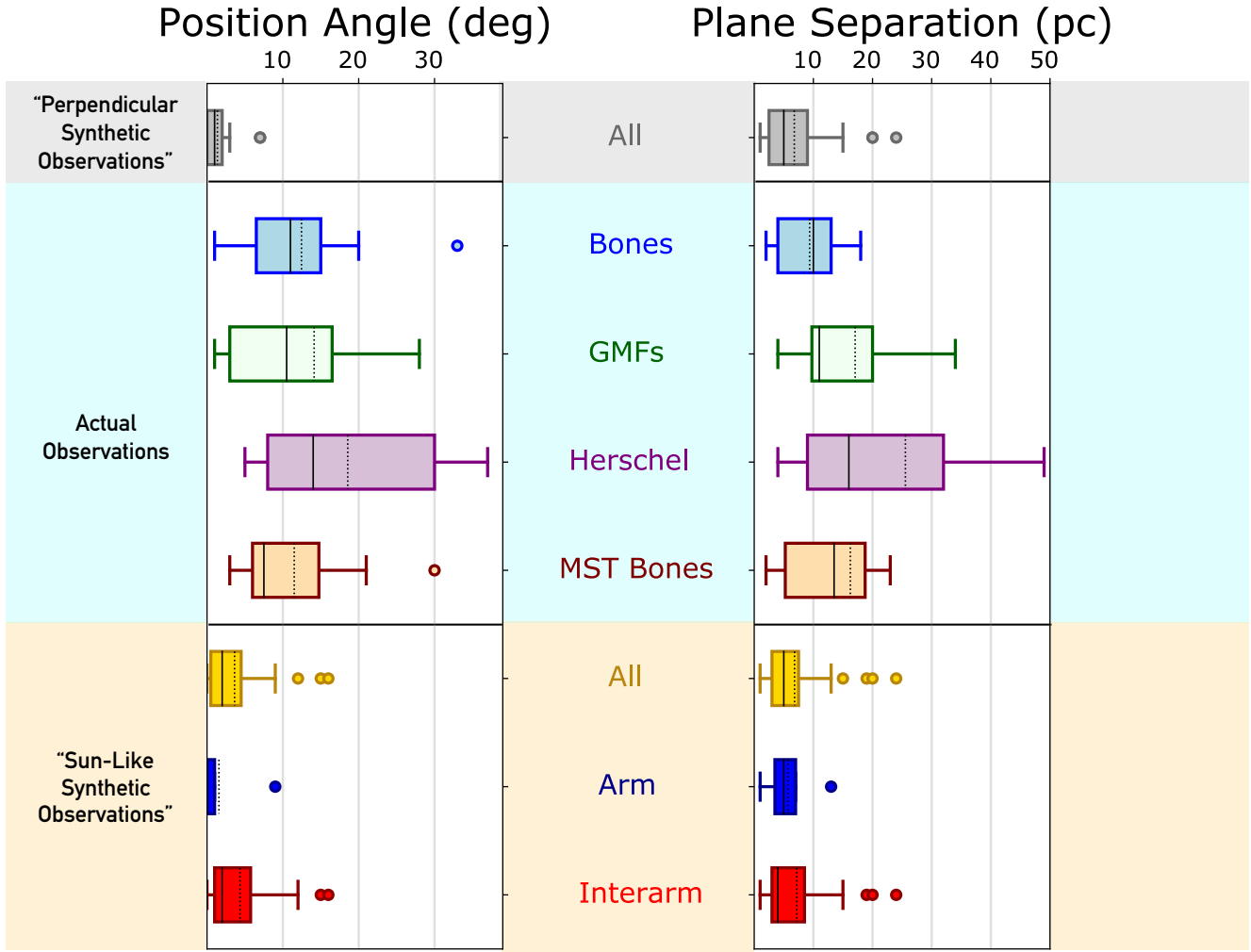}
\caption{ \label{fig:plane}  Same as in Figure \ref{fig:lengths}, but for the distribution of position angle (left) and plane separation (right).}
\end{center}
\end{figure}

Our results for the plane separation and position angle distributions are summarized in Figure \ref{fig:plane}. Both orientation of the observer and galactic environment have no effect on the galactic plane separation, with a typical height of $\approx 5$ pc across all samples. 
We find a similar result for position angle, with the typical position angle across all samples between $\approx 1 - 2^\circ$. 

Most of the observed large-scale Galactic filaments tend to lie near and in close proximity to the physical galactic midplane, typically within 15 pc and oriented less than $30 ^\circ$ from parallel. Several filaments in \citet{Ragan_2014}, \citet{Wang_2015}, and \citet{Wang_2016} catalogs have filaments located $> 60 \; \rm pc$ above the plane. Since a spiral potential is likely the dominant formation mechanism for such elongated structures, it is possible that these filaments were pushed to higher altitudes by an energetic feedback event (e.g. from a nearby supernovae, or HII region). Without feedback in the simulations, it is unsurprising that every filament in the sample forms very close to and aligned with the gravitational midplane of the simulation. This could be the ``natal" state of these filaments prior to the influence of feedback.

While the simulation does not include feedback, it also does not model any warping, flaring, or undulation in the disk, which causes deviation from a flat midplane, particularly in the outer disk \citep[see e.g.][]{Poggio_2018,Momany_2006,Robin_2003, Malhotra_1994}. It is not uncommon to observe molecular clouds at large separations from the Galactic plane \citep[$>$ 100 pc; see e.g. molecular cloud distributions in ][]{Zucker_2019}, so while stellar feedback plays a role, the warping, flaring, and undulation of the disk at larger Galactocentric radii cannot be disregarded in trying to reproduce the scale heights of extended, filamentary, gaseous structures.

\subsection{Velocity Gradients}
\begin{figure}[h!]
\begin{center}
\includegraphics[width=0.75 \columnwidth]{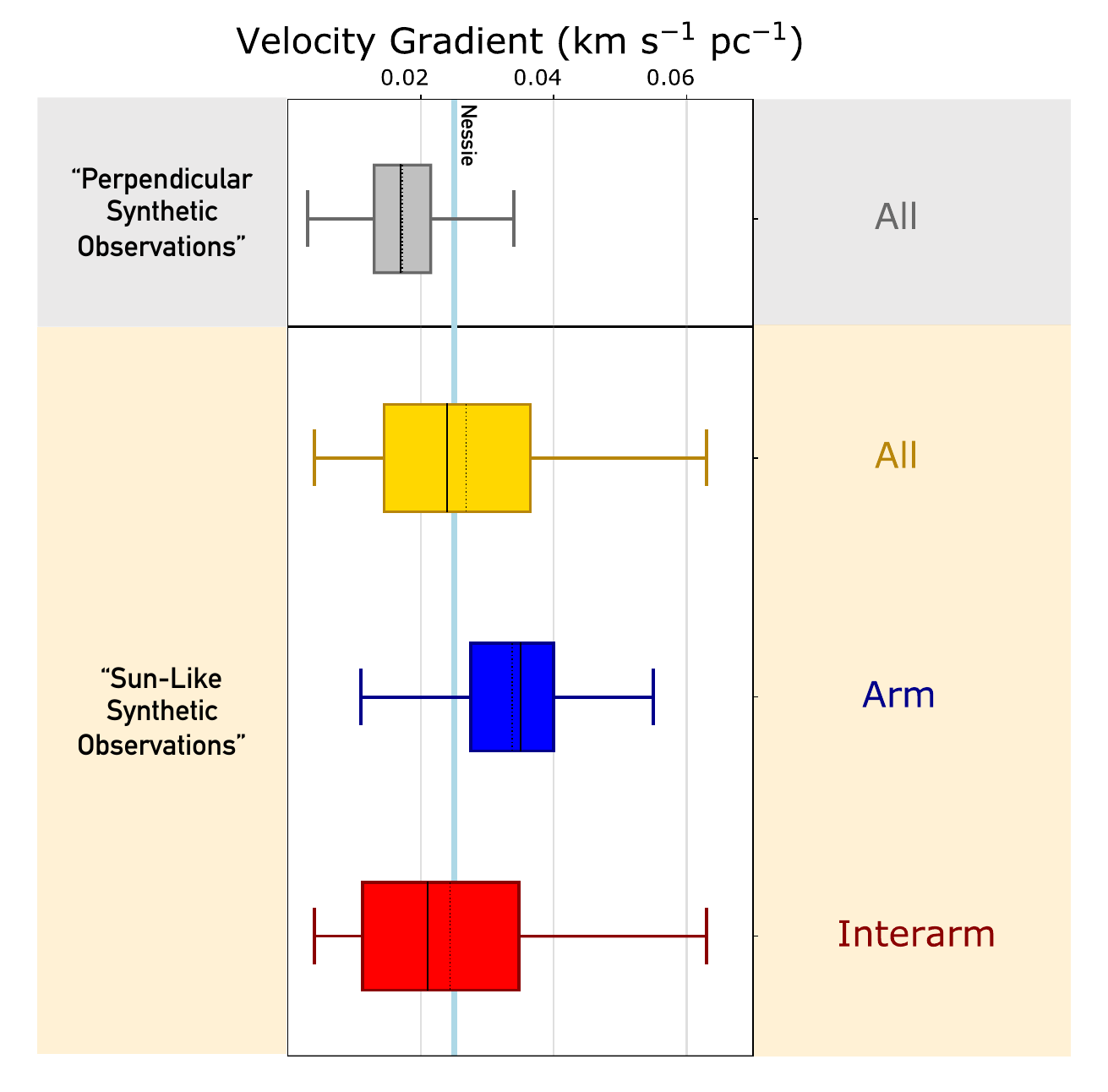}
\caption{ \label{fig:velgrad}  Same as in Figure \ref{fig:lengths}, but for the distribution of velocity gradients along the filaments. The light blue vertical line shows the observed velocity gradient for the paradigmatic Nessie filament \citep{Goodman_2014}}
\end{center}
\end{figure}

To determine the velocity gradients for the filaments, we start by downsampling the filament masks from \S \ref{subsec:identification} by a factor of two, so they can be applied to the $\rm C^{18}O$ spectral cubes (see \ref{subsec:radmc}), whose spatial grids are downgraded by a factor of two for computational expediency before running RADMC-3D. Following the procedure of \citet{Zucker_2018a}, we divide the downsampled masks into bins of width 20 pixels (40 pc). In each bin, we collapse over the pixels within the mask that fall inside the filament mask, and perform a single or multi-component (up to five) Gaussian fit to each bin's spectrum. The Gaussian fitting is performed interactively using the package \texttt{pyspeckit} \citep{pyspeckit}. Since filaments are confirmed to display contiguous \textit{position-velocity} tracks in \S \ref{subsec:identification}, in cases of multiple components, we select the track which aligns with the overall gradient of the filament in \textit{position-velocity} space. Our results are shown in Figure \ref{fig:velgrad}.

We find only modest differences in velocity gradient as a function of both observer position and environment. The typical velocity gradient for filaments in the sun-like projection is $\rm 0.024 \; km \; s^{-1} \; pc^{-1}$, compared to $\rm 0.017 \; km \; s^{-1} \; pc^{-1}$ for the perpendicular projection. The discrepancy between the arm and interarm samples is similar, with velocity gradients in the arm of $\rm 0.035 \; km \; s^{-1} \; pc^{-1}$, compared to $\rm 0.021 \; km \; s^{-1} \; pc^{-1}$ for the interarm regions. In all cases, the velocity gradients are very low, consistent with gradients of only a few kilometers per second over the entire length of the $\approx 100+$ pc long filaments. While a systematic exploration of velocity gradients has not been done for the observed large-scale filament population, the velocity gradients we determine are consistent with those of the paradigmatic ``Nessie" filament. \citet{Goodman_2014} finds a velocity gradient across Nessie of $\rm 0.025 \; km \; s^{-1} \; pc^{-1}$ (or $\approx 4 \; \rm km \; s^{-1}$ over its 160 pc length).

\section{Discussion} \label{sec:discussion}
\subsection{The Origin of Giant Molecular Filaments}

In Figure \ref{fig:cartoon}, we summarize the physical properties of both the synthetic and observed large-scale Galactic filament populations \citep{Zucker_2015, Wang_2015, Wang_2016, Ragan_2014, Abreu_Vicente_2016, Zucker_2018a}. Each filament is shown as a rectangle with the same length, width, galactic plane separation, and position angle as computed uniformly across both the observed and synthetic samples. 

Our results indicate that Giant Molecular Filaments \citep[the ``GMFs", shown in green in Figure \ref{fig:cartoon};][]{Ragan_2014, Abreu_Vicente_2016} appear to be remarkably consistent with the synthetic interarm filament population (shown in red in Figure \ref{fig:cartoon}), without the need to invoke either local self-gravity or stellar feedback. Their lengths, widths, line masses, masses, and column densities all agree with the properties determined for the synthetic interarm filaments, when viewed from the projection of a ``sun-like" observer. This is consistent with the numerical simulations of \citet{Duarte_Cabral_2016, Duarte_Cabral_2017}, who also find association of Giant Molecular Filaments with the interarm regions. Since we are able to reproduce the physical properties of the GMFs using purely galactic dynamics, it is possible that these structures may not be self-gravitating. While the column densities, masses, and linear masses are less consistent overall with GMFs being associated with arms, their lengths and widths are consistent with both the arm and interarm categories, so this does not preclude some GMFs--- particularly those with higher column densities---from being found within spiral arms of the Milky Way. 

The association of observed Giant Molecular Filaments with spiral features in the Milky Way is uncertain, and highly dependent on the study, gas tracer, and set of spiral arm models used. The study of Giant Molecular Filaments by \citet{Ragan_2014} find the seven original GMFs to be preferentially associated with the interarm regions, adopting the spiral arm models of \citet{Vallee_2008} and using primarily $\rm ^{13}CO$ data from GRS \citep{Jackson_2006} to confirm velocity contiguity. However, an expanded follow-up study by \citet{Abreu_Vicente_2016} -- using the updated spiral arm models from \citet{Reid_2016} and $\rm ^{13}CO$ from ThrUMMS \citep{Barnes_2015} -- find nine of sixteen GMFs to be associated with arms. The most recent study by \citet{Zucker_2018a}, which uses the spiral arm models from \citet{Reid_2016} to determine arm association of large-scale filaments in the first quadrant, find that only one third of GMFs show strong association with spiral arms in both $l-b$ and $p-v$ space. However, \citet{Zucker_2018a} mainly consider $\rm ^{13}CO$ data (with ancillary dense gas information), and only for a subset of the sample. More broadly, the large-scale filament identification methods from previous work applied different velocity information, with some studies relying only on dense gas tracers \citep{Wang_2016}, others using only CO \citep{Wang_2015} and others primarily considering CO but supplementing with dense gas tracers \citep{Ragan_2014, Abreu_Vicente_2016, Zucker_2015, Zucker_2018a}. Given the wide variety of kinematic tracers used through the large-scale filament literature in the past, a systematic study of the kinematics using consistent methodologies and spectral-line surveys across the full sample is needed to better constrain the relationship between GMFs (and the greater population of large-scale Galactic filaments) and spiral arms.

\subsection{The Origin of Nessie} 
\begin{figure}[h!]
\begin{center}
\includegraphics[width=1. \columnwidth]{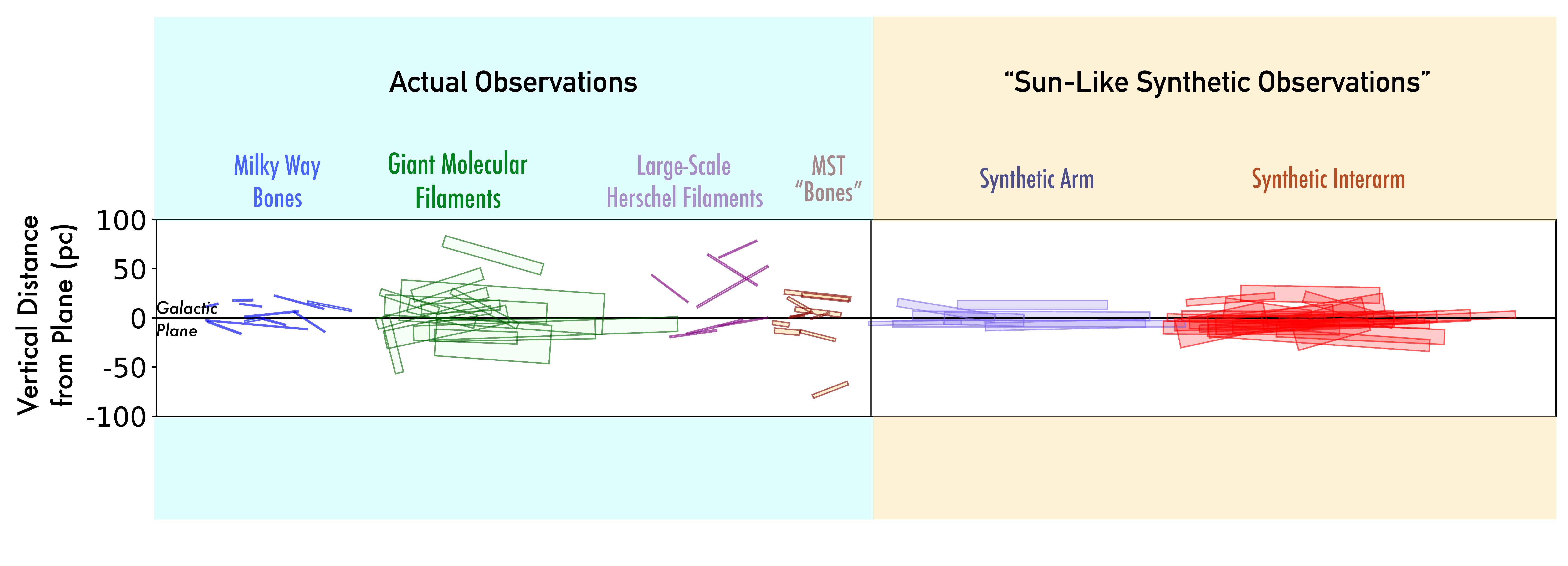}
\caption{ \label{fig:cartoon}  Cartoon illustrating the differences between the observed large scale filament population (left panel) and the synthetic large scale filament population lying in the arm and interarm regions (right panel), as seen in the ``sun-like" projection. Each filament is represented as a rectangle with the same length, width, galactic plane separation, and position angle as computed uniformly across both the observed and synthetic samples. }
\end{center}
\end{figure}

Unlike the Giant Molecular Filaments, we are unable to reproduce the properties of 
``Bone-like" filaments similar to Nessie \citep{Goodman_2014, Zucker_2015, Zucker_2018a} (the light blue population in Figure \ref{fig:cartoon}) in simulations which only include a spiral potential. As a result, the formation of ``Bone-like" filaments likely require a Galactic potential in combination with both feedback and local self-gravity. Nessie is anywhere from 160-430 pc long \citep[depending on the strictness with which one connects different parts of Nessie disrupted by feedback;][]{Goodman_2014} and has a width of $\approx 1$ pc seen in dust emission. Other dense ``Bone-like" filaments, whose lengths are likely more subject to projection effects \citep[see][]{Zucker_2018a} have lengths on the order of $\approx 50$ pc and similar widths. Thus, on average, the observed ``Bone-like" filament population has a factor of ten smaller width, and a factor of a few smaller length when compared to the synthetic population presented in this work. Observed Bone samples \citep{Goodman_2014, Zucker_2015, Zucker_2018a} show strong agreement with spiral arms models in \textit{position-position-velocity} space. However, while the column density of Nessie is significantly higher than the GMFs, the observed column densities of the Bones from \citet{Zucker_2019} are still lower than the synthetic arm filaments. Recall that the  \citet{Zucker_2019} column densities for the Bones are beam-diluted, given that the typical resolution of Herschel ($\approx 1$ pc at 3 kpc) is equivalent to the width of Nessie. Assuming Nessie dominates the column density along the line of sight, higher resolution column density studies of the filament --- using a combination of VVV \citep{Saito_2012} and Spitzer/GLIMPSE \citep{Churchwell_2009} data --- determine a typical $A_V$ along Nessie of 20 mag, or $\rm N_{H_2} \approx 2\times 10^{22} \; cm^{-2}$, given standard conversion factors, in better agreement with the fiducial column density of $\rm 4\times 10^{22} \; cm^{-2}$ for the synthetic arm filaments \citep[c.f.][]{Mattern_2018a}. 

Nevertheless, because the lengths of synthetic large-scale filaments in both environments are too long without feedback, and their widths too puffy without local self-gravity, it is difficult to definitively determine which population (arm or interarm, or both) is most consistent with Nessie-like structures. It is clear that while a spiral potential is required to form Nessie-like filaments, both feedback and local self-gravity likely also play an important role in shaping their physical properties. 

Smith et al. (2019, in prep), presents an updated treatment of the \citet{Smith_2014} simulations utilized in this work, investigating how the incorporation of additional physics (including local self-gravity, random supernovae feedback, and feedback tied to sink particles) influences cloud properties (at $\approx 0.1$ pc resolution) in concert with a spiral potential, over a smaller fraction of the disk. While Smith et al. 2019 (in prep) explores the effects of feedback and self-gravity for smaller scale filaments (lengths $\approx$ a few parsecs, widths $\approx$ a few tenths of a parsec), we plan to revisit the effect of this physics, to better characterize the relative influence of galactic dynamics, feedback, and local self-gravity on large-scale filaments coincident with both the arm and interarm regions.

\subsection{Current Galactic Environment Drives Filament Properties}
We find that, overall, current galactic environment (``arm" vs. ``interarm"), rather than projection effects, plays the dominant role in setting the physical properties of the filaments. Mass, linear mass, width, galactic plane separation, and position angle (the filament's 2D projected orientation with respect to the midplane) are largely invariant to projection effects. The length is the most sensitive to projection effects, with the filament lengths we observe from the ``sun-like" projection typically shortened by a factor of $1.4$ compared to the intrinsic lengths of the filaments (seen in the ``perpendicular" projection). However, this can be as high as a factor of three to four for extreme observer viewing angles. Column density mildly increases due to projection effects (by $\approx 30\%$), but the magnitude of the effect we would actually observe is more uncertain, given that we only consider the column density locally around the filament, and do not consider intervening column density along the line of sight between the ``sun" and each filament.


In contrast, galactic environment plays a significant role in setting the column densities of the filaments, with those in the arm exhibiting typical column densities $\approx 7\times$ higher than in the interarm regions. This propagates to the inferred masses and linear masses, with again, these properties being significantly ($\approx 3-5 \times$) higher in the arm. All of these differences naturally follow from the idea that filaments in the arms are subject to higher pressures as they enter the spiral-potential wells.

One property that appears invariant to galactic environment is the length, which is similar in both the arm and interarm regions ($\approx 150$ pc). It is possible that the inclusion of additional physics (particularly feedback) could induce larger variations in length than those seen here. Since most feedback should be confined to  spiral arms \citep{Bartunov_1994, Aramyan_2016}, filaments in the arms are more likely to be broken apart by feedback events. They are also more likely to be merge and interact with other clouds, and be subject to cloud-cloud confusion, which could limit their identification on hundred parsec scales \citep{Duarte_Cabral_2016, Duarte_Cabral_2017}. Regardless, the inclusion of both local self-gravity and feedback will have a fundamental effect on the morphology of dense gas. Our results should be treated in the context of the properties being upper and lower limits (as stated in Table \ref{tab:caution}), until we are able to explore these effects over smaller areas of the disc in the more advanced numerical simulations presented in Smith et al. (2019, in prep). 

\section{Conclusion} \label{sec:conclusion}

The \citet{Smith_2014} simulations present an opportunity to analyze the response of a gaseous disk to an imposed analytic potential in the context of a simple $\rm H_2$ and CO chemical model. Its large dynamic range ($<$ 1 pc resolution in regions with $\rm n > 10^{3} \; cm^{-3}$) over a large area of the disk ($\rm \approx 30 \; kpc^{2}$) is ideal for resolving both the widths of the large-scale filaments and the galactic environments in which they form. We analyze the physical properties of synthetic large-scale filaments as a function of both projection effects and galactic environment, and compare them to observations. Our conclusions are as follows:

\begin{itemize}
    \item Observer viewing angle has a mild effect on filament properties, with filament widths, masses, plane separations, and position angles varying by $\lesssim 25\%$ due to the position of the observer in the Galaxy. The length is most influenced by projection effects, with the typical filament length foreshortened by a factor of $\approx 1.5\times$ in comparison to the idealized ``perpendicular" projection. In some cases, however, when the inclination angle between the observer and the long-axis of the filament is very high, the length can be foreshortened by $3\times$.
    \item There is significant variation in filament properties based on current galactic environment (arm vs. interarm). This follows naturally from filaments in the arms likely being subject to higher pressures as they enter the spiral-potential wells compared to the interarm regions, which may form via differential rotation. The column densities of filaments in the arms are typically $7 \times$ higher than in the interarm regions, and their masses and linear masses are $\approx 3-5 \times$ higher.
    \item In this simplified, spiral potential only case, we are able to broadly reproduce the lengths, widths, column densities, masses, and linear masses of the ``Giant Molecular Filaments" \citep{Ragan_2014, Abreu_Vicente_2016, Zhang_2019} using the synthetic ``interarm" filament sample observed from the position of a sun-based observer. We cannot reproduce the the physical properties of ``Bone-like" filaments like Nessie, as the widths are a factor of $10\times$ too large without local self-gravity, and on average $2-3\times$ too long without stellar feedback. 
    \item We estimate that we should be able to identify no more than $\approx 1000$ large-scale filaments across the whole Milky Way. These filaments should be destroyed very quickly (within a few million years) due to internal feedback from HII regions or supernovae, along with cloud-cloud collisions in the arms. Nevertheless, they are found in ``special" places in the galaxy, within the gravitational midplane, and tracing either the spiral-potential wells, or dynamic interarm features which develop in response to a spiral potential. A time-dependent analysis of the filaments' evolution \citep[c.f.][]{Duarte_Cabral_2017} is needed to further constrain potential formation mechanisms. More advanced numerical simulations, which incorporate feedback and local self-gravity (see the Smith et al. 2019 ``Cloud Factory" simulations) will also allow us to analyze these effects in more detail, and better contextualize the properties of synthetic large-scale filaments in light of the diverse populations of hundred-parsec-scale filaments observed in our own Galaxy.

\end{itemize}

\software{
 \texttt{astropy} \citep{Astropy_2018},\texttt{glue} \citep{glueviz_2017}, \texttt{yt}, \citep{Turk_2011} \texttt{RADMC-3D} \citep{Dullemond_2012}, \texttt{matplotlib} \citep{Hunter_2007}, \texttt{numpy} \citep{numpy}} \\

\acknowledgements
We thank an anonymous referee for their feedback, the implementation of which improved the quality of this manuscript. \\

The authors acknowledge Paris-Saclay University's Institut Pascal program ``The Self-Organized Star Formation Process" and the Interstellar Institute for hosting discussions that nourished the development of the ideas behind this work.\\

C.Z. is supported by the NSF Graduate Research Fellowship Program (Grant No. 1650114) and the Harvard Data Science Initiative. \\

RJS gratefully acknowledges support from an STFC Ernest Rutherford Fellowship (grant ST/N00485X/1). This work used the COSMA Data Centric system at Durham University, operated by the Institute for Computational Cosmology on behalf of the STFC DiRAC HPC Facility (www.dirac.ac.uk. This equipment was funded by a BIS National E-infrastructure capital grant ST/K00042X/1, DiRAC Operations grant ST/K003267/1 and Durham University. \\

The visualization, exploration, and interpretation of data presented in this work was made possible using the glue visualization software, supported under NSF grant OAC-1739657. \\

\bibliography{full_article}

\begin{thebibliography}{}
\expandafter\ifx\csname natexlab\endcsname\relax\def\natexlab#1{#1}\fi
\providecommand{\url}[1]{\href{#1}{#1}}
\providecommand{\dodoi}[1]{doi:~\href{http://doi.org/#1}{\nolinkurl{#1}}}
\providecommand{\doeprint}[1]{\href{http://ascl.net/#1}{\nolinkurl{http://ascl.net/#1}}}
\providecommand{\doarXiv}[1]{\href{https://arxiv.org/abs/#1}{\nolinkurl{https://arxiv.org/abs/#1}}}

\bibitem[{Abreu-Vicente {et~al.}(2016)Abreu-Vicente, Ragan, Kainulainen,
  Henning, Beuther, \& Johnston}]{Abreu_Vicente_2016}
Abreu-Vicente, J., Ragan, S., Kainulainen, J., {et~al.} 2016, Astronomy {\&}
  Astrophysics, 590, A131, \dodoi{10.1051/0004-6361/201527674}

\bibitem[{{Andr{\'e}} {et~al.}(2010){Andr{\'e}}, {Men'shchikov}, {Bontemps},
  {K{\"o}nyves}, {Motte}, {Schneider}, {Didelon}, {Minier}, {Saraceno},
  {Ward-Thompson}, {di Francesco}, {White}, {Molinari}, {Testi}, {Abergel},
  {Griffin}, {Henning}, {Royer}, {Mer{\'\i}n}, {Vavrek}, {Attard},
  {Arzoumanian}, {Wilson}, {Ade}, {Aussel}, {Baluteau}, {Benedettini},
  {Bernard}, {Blommaert}, {Cambr{\'e}sy}, {Cox}, {di Giorgio}, {Hargrave},
  {Hennemann}, {Huang}, {Kirk}, {Krause}, {Launhardt}, {Leeks}, {Le Pennec},
  {Li}, {Martin}, {Maury}, {Olofsson}, {Omont}, {Peretto}, {Pezzuto}, {Prusti},
  {Roussel}, {Russeil}, {Sauvage}, {Sibthorpe}, {Sicilia-Aguilar}, {Spinoglio},
  {Waelkens}, {Woodcraft}, \& {Zavagno}}]{Andre_2010}
{Andr{\'e}}, P., {Men'shchikov}, A., {Bontemps}, S., {et~al.} 2010, \aap, 518,
  L102, \dodoi{10.1051/0004-6361/201014666}

\bibitem[{{Aramyan} {et~al.}(2016){Aramyan}, {Hakobyan}, {Petrosian}, {de
  Lapparent}, {Bertin}, {Mamon}, {Kunth}, {Nazaryan}, {Adibekyan}, \&
  {Turatto}}]{Aramyan_2016}
{Aramyan}, L.~S., {Hakobyan}, A.~A., {Petrosian}, A.~R., {et~al.} 2016, \mnras,
  459, 3130, \dodoi{10.1093/mnras/stw873}

\bibitem[{Arzoumanian {et~al.}(2011)Arzoumanian, Andr{\'{e}}, Didelon,
  KÃ¶nyves, Schneider, Men'shchikov, Sousbie, Zavagno, Bontemps, Francesco,
  Griffin, Hennemann, Hill, Kirk, Martin, Minier, Molinari, Motte, Peretto,
  Pezzuto, Spinoglio, Ward-Thompson, White, \& Wilson}]{Arzoumanian_2011}
Arzoumanian, D., Andr{\'{e}}, P., Didelon, P., {et~al.} 2011, Astronomy {\&}
  Astrophysics, 529, L6, \dodoi{10.1051/0004-6361/201116596}

\bibitem[{{Astropy Collaboration} {et~al.}(2018){Astropy Collaboration},
  {Price-Whelan}, {Sip{\'{o}}cz}, {G{\"u}nther}, {Lim}, {Crawford}, {Conseil},
  {Shupe}, {Craig}, {Dencheva}, {Ginsburg}, {VanderPlas}, {Bradley},
  {P{\'e}rez-Su{\'a}rez}, {de Val- Borro}, {Aldcroft}, {Cruz}, {Robitaille},
  {Tollerud}, {Ardelean}, {Babej}, {Bach}, {Bachetti}, {Bakanov}, {Bamford},
  {Barentsen}, {Barmby}, {Baumbach}, {Berry}, {Biscani}, {Boquien}, {Bostroem},
  {Bouma}, {Brammer}, {Bray}, {Breytenbach}, {Buddelmeijer}, {Burke},
  {Calderone}, {Cano Rodr{\'\i}guez}, {Cara}, {Cardoso}, {Cheedella}, {Copin},
  {Corrales}, {Crichton}, {D'Avella}, {Deil}, {Depagne}, {Dietrich}, {Donath},
  {Droettboom}, {Earl}, {Erben}, {Fabbro}, {Ferreira}, {Finethy}, {Fox},
  {Garrison}, {Gibbons}, {Goldstein}, {Gommers}, {Greco}, {Greenfield},
  {Groener}, {Grollier}, {Hagen}, {Hirst}, {Homeier}, {Horton}, {Hosseinzadeh},
  {Hu}, {Hunkeler}, {Ivezi{\'c}}, {Jain}, {Jenness}, {Kanarek}, {Kendrew},
  {Kern}, {Kerzendorf}, {Khvalko}, {King}, {Kirkby}, {Kulkarni}, {Kumar},
  {Lee}, {Lenz}, {Littlefair}, {Ma}, {Macleod}, {Mastropietro}, {McCully},
  {Montagnac}, {Morris}, {Mueller}, {Mumford}, {Muna}, {Murphy}, {Nelson},
  {Nguyen}, {Ninan}, {N{\"o}the}, {Ogaz}, {Oh}, {Parejko}, {Parley}, {Pascual},
  {Patil}, {Patil}, {Plunkett}, {Prochaska}, {Rastogi}, {Reddy Janga},
  {Sabater}, {Sakurikar}, {Seifert}, {Sherbert}, {Sherwood-Taylor}, {Shih},
  {Sick}, {Silbiger}, {Singanamalla}, {Singer}, {Sladen}, {Sooley},
  {Sornarajah}, {Streicher}, {Teuben}, {Thomas}, {Tremblay}, {Turner},
  {Terr{\'o}n}, {van Kerkwijk}, {de la Vega}, {Watkins}, {Weaver}, {Whitmore},
  {Woillez}, {Zabalza}, \& {Astropy Contributors}}]{Astropy_2018}
{Astropy Collaboration}, {Price-Whelan}, A.~M., {Sip{\'{o}}cz}, B.~M., {et~al.}
  2018, \aj, 156, 123, \dodoi{10.3847/1538-3881/aabc4f}

\bibitem[{Ballesteros-Paredes {et~al.}(2011)Ballesteros-Paredes,
  V{\'{a}}zquez-Semadeni, Gazol, Hartmann, Heitsch, \&
  Col{\'{\i}}n}]{Ballesteros_Paredes_2011}
Ballesteros-Paredes, J., V{\'{a}}zquez-Semadeni, E., Gazol, A., {et~al.} 2011,
  Monthly Notices of the Royal Astronomical Society, 416, 1436,
  \dodoi{10.1111/j.1365-2966.2011.19141.x}

\bibitem[{Barnes {et~al.}(2015)Barnes, Muller, Indermuehle, O'Dougherty, Lowe,
  Cunningham, Hernandez, \& Fuller}]{Barnes_2015}
Barnes, P.~J., Muller, E., Indermuehle, B., {et~al.} 2015, {ApJ}, 812, 6,
  \dodoi{10.1088/0004-637x/812/1/6}

\bibitem[{{Bartunov} {et~al.}(1994){Bartunov}, {Tsvetkov}, \&
  {Filimonova}}]{Bartunov_1994}
{Bartunov}, O.~S., {Tsvetkov}, D.~Y., \& {Filimonova}, I.~V. 1994, Publications
  of the Astronomical Society of the Pacific, 106, 1276, \dodoi{10.1086/133505}

\bibitem[{{Benjamin} {et~al.}(2005){Benjamin}, {Churchwell}, {Babler},
  {Indebetouw}, {Meade}, {Whitney}, {Watson}, {Wolfire}, {Wolff}, {Ignace},
  {Bania}, {Bracker}, {Clemens}, {Chomiuk}, {Cohen}, {Dickey}, {Jackson},
  {Kobulnicky}, {Mercer}, {Mathis}, {Stolovy}, \& {Uzpen}}]{Benjamin_2005}
{Benjamin}, R.~A., {Churchwell}, E., {Babler}, B.~L., {et~al.} 2005, \apj, 630,
  L149, \dodoi{10.1086/491785}

\bibitem[{{Bonnell} {et~al.}(2006){Bonnell}, {Dobbs}, {Robitaille}, \&
  {Pringle}}]{Bonnell_2006}
{Bonnell}, I.~A., {Dobbs}, C.~L., {Robitaille}, T.~P., \& {Pringle}, J.~E.
  2006, \mnras, 365, 37, \dodoi{10.1111/j.1365-2966.2005.09657.x}

\bibitem[{{Burkhart} {et~al.}(2017){Burkhart}, {Stalpes}, \&
  {Collins}}]{Burkhart_2017}
{Burkhart}, B., {Stalpes}, K., \& {Collins}, D.~C. 2017, \apjl, 834, L1,
  \dodoi{10.3847/2041-8213/834/1/L1}

\bibitem[{Carey {et~al.}(2009)Carey, Noriega-Crespo, Mizuno, Shenoy, Paladini,
  Kraemer, Price, Flagey, Ryan, Ingalls, Kuchar, Gon{\c{c}}alves, Indebetouw,
  Billot, Marleau, Padgett, Rebull, Bressert, Ali, Molinari, Martin, Berriman,
  Boulanger, Latter, Miville-Deschenes, Shipman, \& Testi}]{Carey_2009}
Carey, S.~J., Noriega-Crespo, A., Mizuno, D.~R., {et~al.} 2009, {PUBL} {ASTRON}
  {SOC} {PAC}, 121, 76, \dodoi{10.1086/596581}

\bibitem[{{Churchwell} {et~al.}(2009){Churchwell}, {Babler}, {Meade},
  {Whitney}, {Benjamin}, {Indebetouw}, {Cyganowski}, {Robitaille}, {Povich},
  {Watson}, \& {Bracker}}]{Churchwell_2009}
{Churchwell}, E., {Babler}, B.~L., {Meade}, M.~R., {et~al.} 2009, \pasp, 121,
  213, \dodoi{10.1086/597811}

\bibitem[{{Clark} {et~al.}(2012){Clark}, {Glover}, {Klessen}, \&
  {Bonnell}}]{Clark_2012}
{Clark}, P.~C., {Glover}, S. C.~O., {Klessen}, R.~S., \& {Bonnell}, I.~A. 2012,
  \mnras, 424, 2599, \dodoi{10.1111/j.1365-2966.2012.21259.x}

\bibitem[{{Colombo} {et~al.}(2015){Colombo}, {Rosolowsky}, {Ginsburg},
  {Duarte-Cabral}, \& {Hughes}}]{Colombo_2015}
{Colombo}, D., {Rosolowsky}, E., {Ginsburg}, A., {Duarte-Cabral}, A., \&
  {Hughes}, A. 2015, \mnras, 454, 2067, \dodoi{10.1093/mnras/stv2063}

\bibitem[{{Colombo} {et~al.}(2019){Colombo}, {Rosolowsky}, {Duarte-Cabral},
  {Ginsburg}, {Glenn}, {Zetterlund}, {Hernandez}, {Dempsey}, \&
  {Currie}}]{Colombo_2019}
{Colombo}, D., {Rosolowsky}, E., {Duarte-Cabral}, A., {et~al.} 2019, \mnras,
  483, 4291, \dodoi{10.1093/mnras/sty3283}

\bibitem[{{Dobbs} \& {Bonnell}(2006)}]{Dobbs_2006}
{Dobbs}, C.~L., \& {Bonnell}, I.~A. 2006, \mnras, 367, 873,
  \dodoi{10.1111/j.1365-2966.2006.10146.x}

\bibitem[{{Dobbs} \& {Bonnell}(2008)}]{Dobbs_2008}
---. 2008, \mnras, 385, 1893, \dodoi{10.1111/j.1365-2966.2008.12995.x}

\bibitem[{{Du} {et~al.}(2017){Du}, {Xu}, {Yang}, \& {Sun}}]{Du_2017}
{Du}, X., {Xu}, Y., {Yang}, J., \& {Sun}, Y. 2017, \apjs, 229, 24,
  \dodoi{10.3847/1538-4365/aa5d9d}

\bibitem[{Duarte-Cabral \& Dobbs(2016)}]{Duarte_Cabral_2016}
Duarte-Cabral, A., \& Dobbs, C.~L. 2016, Mon. Not. R. Astron. Soc., 458, 3667,
  \dodoi{10.1093/mnras/stw469}

\bibitem[{{Duarte-Cabral} \& {Dobbs}(2017)}]{Duarte_Cabral_2017}
{Duarte-Cabral}, A., \& {Dobbs}, C.~L. 2017, ArXiv e-prints.
\newblock \doarXiv{1706.05421}

\bibitem[{{Dullemond} {et~al.}(2012){Dullemond}, {Juhasz}, {Pohl}, {Sereshti},
  {Shetty}, {Peters}, {Commercon}, \& {Flock}}]{Dullemond_2012}
{Dullemond}, C.~P., {Juhasz}, A., {Pohl}, A., {et~al.} 2012, {RADMC-3D: A
  multi-purpose radiative transfer tool}.
\newblock \doeprint{1202.015}

\bibitem[{{Ginsburg} \& {Mirocha}(2011)}]{pyspeckit}
{Ginsburg}, A., \& {Mirocha}, J. 2011, {PySpecKit: Python Spectroscopic
  Toolkit}.
\newblock \doeprint{1109.001}

\bibitem[{{Glover} \& {Mac Low}(2007{\natexlab{a}})}]{Glover_2007a}
{Glover}, S. C.~O., \& {Mac Low}, M.-M. 2007{\natexlab{a}}, The Astrophysical
  Journal Supplement Series, 169, 239, \dodoi{10.1086/512238}

\bibitem[{{Glover} \& {Mac Low}(2007{\natexlab{b}})}]{Glover_2007b}
---. 2007{\natexlab{b}}, \apj, 659, 1317, \dodoi{10.1086/512227}

\bibitem[{Goodman {et~al.}(2014)Goodman, Alves, Beaumont, Benjamin, Borkin,
  Burkert, Dame, Jackson, Kauffmann, Robitaille, \& Smith}]{Goodman_2014}
Goodman, A.~A., Alves, J., Beaumont, C.~N., {et~al.} 2014, {ApJ}, 797, 53,
  \dodoi{10.1088/0004-637x/797/1/53}

\bibitem[{{Hacar} {et~al.}(2018){Hacar}, {Tafalla}, {Forbrich}, {Alves},
  {Meingast}, {Grossschedl}, \& {Teixeira}}]{Hacar_2018}
{Hacar}, A., {Tafalla}, M., {Forbrich}, J., {et~al.} 2018, \aap, 610, A77,
  \dodoi{10.1051/0004-6361/201731894}

\bibitem[{Hacar {et~al.}(2013)Hacar, Tafalla, Kauffmann, \&
  Kov{\'{a}}cs}]{Hacar_2013}
Hacar, A., Tafalla, M., Kauffmann, J., \& Kov{\'{a}}cs, A. 2013, Astronomy {\&}
  Astrophysics, 554, A55, \dodoi{10.1051/0004-6361/201220090}

\bibitem[{{Heyer} \& {Brunt}(2004)}]{Heyer_2004}
{Heyer}, M.~H., \& {Brunt}, C.~M. 2004, \apj, 615, L45, \dodoi{10.1086/425978}

\bibitem[{Hunter(2007)}]{Hunter_2007}
Hunter, J.~D. 2007, Computing in Science \& Engineering, 9, 90,
  \dodoi{10.1109/MCSE.2007.55}

\bibitem[{Jackson {et~al.}(2010)Jackson, Finn, Chambers, Rathborne, \&
  Simon}]{Jackson_2010}
Jackson, J.~M., Finn, S.~C., Chambers, E.~T., Rathborne, J.~M., \& Simon, R.
  2010, {ApJ}, 719, L185, \dodoi{10.1088/2041-8205/719/2/l185}

\bibitem[{Jackson {et~al.}(2006)Jackson, Rathborne, Shah, Simon, Bania,
  Clemens, Chambers, Johnson, Dormody, Lavoie, \& Heyer}]{Jackson_2006}
Jackson, J.~M., Rathborne, J.~M., Shah, R.~Y., {et~al.} 2006, The Astrophysical
  Journal Supplement Series, 163, 145, \dodoi{10.1086/500091}

\bibitem[{Juvela {et~al.}(2012)Juvela, Ristorcelli, Pagani, Doi, Pelkonen,
  Marshall, Bernard, Falgarone, Malinen, Marton, McGehee, Montier, Motte,
  Paladini, T{\'{o}}th, Ysard, Zahorecz, \& Zavagno}]{Juvela_2012}
Juvela, M., Ristorcelli, I., Pagani, L., {et~al.} 2012, Astronomy {\&}
  Astrophysics, 541, A12, \dodoi{10.1051/0004-6361/201118640}

\bibitem[{Kauffmann {et~al.}(2008)Kauffmann, Bertoldi, Bourke, Evans, \&
  Lee}]{Kauffmann_2008}
Kauffmann, J., Bertoldi, F., Bourke, T.~L., Evans, N.~J., \& Lee, C.~W. 2008,
  Astronomy and Astrophysics, 487, 993, \dodoi{10.1051/0004-6361:200809481}

\bibitem[{{Kim} \& {Ostriker}(2002)}]{Kim_2002}
{Kim}, W.-T., \& {Ostriker}, E.~C. 2002, \apj, 570, 132, \dodoi{10.1086/339352}

\bibitem[{Koch \& Rosolowsky(2015)}]{Koch_2015}
Koch, E.~W., \& Rosolowsky, E.~W. 2015, Mon. Not. R. Astron. Soc., 452, 3435,
  \dodoi{10.1093/mnras/stv1521}

\bibitem[{Li {et~al.}(2016)Li, Urquhart, Leurini, Csengeri, Wyrowski, Menten,
  \& Schuller}]{Li_2016}
Li, G.-X., Urquhart, J.~S., Leurini, S., {et~al.} 2016, Astronomy {\&}
  Astrophysics, 591, A5, \dodoi{10.1051/0004-6361/201527468}

\bibitem[{Li {et~al.}(2013)Li, Wyrowski, Menten, \& Belloche}]{Li_2013}
Li, G.-X., Wyrowski, F., Menten, K., \& Belloche, A. 2013, Astronomy {\&}
  Astrophysics, 559, A34, \dodoi{10.1051/0004-6361/201322411}

\bibitem[{{Malhotra}(1994)}]{Malhotra_1994}
{Malhotra}, S. 1994, \apj, 433, 687, \dodoi{10.1086/174677}

\bibitem[{{Mattern} {et~al.}(2018{\natexlab{a}}){Mattern}, {Kainulainen},
  {Zhang}, \& {Beuther}}]{Mattern_2018a}
{Mattern}, M., {Kainulainen}, J., {Zhang}, M., \& {Beuther}, H.
  2018{\natexlab{a}}, \aap, 616, A78, \dodoi{10.1051/0004-6361/201731778}

\bibitem[{{Mattern} {et~al.}(2018{\natexlab{b}}){Mattern}, {Kauffmann},
  {Csengeri}, {Urquhart}, {Leurini}, {Wyrowski}, {Giannetti}, {Barnes},
  {Beuther}, {Bronfman}, {Duarte-Cabral}, {Henning}, {Kainulainen}, {Menten},
  {Schisano}, \& {Schuller}}]{Mattern_2018b}
{Mattern}, M., {Kauffmann}, J., {Csengeri}, T., {et~al.} 2018{\natexlab{b}},
  \aap, 619, A166, \dodoi{10.1051/0004-6361/201833406}

\bibitem[{{Miville-Desch{\^e}nes} {et~al.}(2017){Miville-Desch{\^e}nes},
  {Murray}, \& {Lee}}]{Miville_Deschenes_2017}
{Miville-Desch{\^e}nes}, M.-A., {Murray}, N., \& {Lee}, E.~J. 2017, \apj, 834,
  57, \dodoi{10.3847/1538-4357/834/1/57}

\bibitem[{Molinari {et~al.}(2016)Molinari, Schisano, Elia, Pestalozzi,
  Traficante, Pezzuto, Swinyard, Noriega-Crespo, Bally, Moore, Plume, Zavagno,
  di~Giorgio, Liu, Pilbratt, Mottram, Russeil, Piazzo, Veneziani, Benedettini,
  Calzoletti, Faustini, Natoli, Piacentini, Merello, Palmese, Grande,
  Polychroni, Rygl, Polenta, Barlow, Bernard, Martin, Testi, Ali, Andr{\'{e}},
  Beltr{\'{a}}n, Billot, Carey, Cesaroni, Compi{\`{e}}gne, Eden, Fukui,
  Garcia-Lario, Hoare, Huang, Joncas, Lim, Lord, Martinavarro-Armengol, Motte,
  Paladini, Paradis, Peretto, Robitaille, Schilke, Schneider, Schulz,
  Sibthorpe, Strafella, Thompson, Umana, Ward-Thompson, \&
  Wyrowski}]{Molinari_2016}
Molinari, S., Schisano, E., Elia, D., {et~al.} 2016, Astronomy {\&}
  Astrophysics, 591, A149, \dodoi{10.1051/0004-6361/201526380}

\bibitem[{{Momany} {et~al.}(2006){Momany}, {Zaggia}, {Gilmore}, {Piotto},
  {Carraro}, {Bedin}, \& {de Angeli}}]{Momany_2006}
{Momany}, Y., {Zaggia}, S., {Gilmore}, G., {et~al.} 2006, \aap, 451, 515,
  \dodoi{10.1051/0004-6361:20054081}

\bibitem[{{Nelson} \& {Langer}(1997)}]{Nelson_1997}
{Nelson}, R.~P., \& {Langer}, W.~D. 1997, \apj, 482, 796,
  \dodoi{10.1086/304167}

\bibitem[{{Ossenkopf}(1997)}]{Ossenkopf_1997}
{Ossenkopf}, V. 1997, \na, 2, 365, \dodoi{10.1016/S1384-1076(97)00026-2}

\bibitem[{{Poggio} {et~al.}(2018){Poggio}, {Drimmel}, {Lattanzi}, {Smart},
  {Spagna}, {Andrae}, {Bailer-Jones}, {Fouesneau}, {Antoja}, {Babusiaux},
  {Evans}, {Figueras}, {Katz}, {Reyl{\'e}}, {Robin}, {Romero-G{\'o}mez}, \&
  {Seabroke}}]{Poggio_2018}
{Poggio}, E., {Drimmel}, R., {Lattanzi}, M.~G., {et~al.} 2018, \mnras, 481,
  L21, \dodoi{10.1093/mnrasl/sly148}

\bibitem[{Ragan {et~al.}(2014)Ragan, Henning, Tackenberg, Beuther, Johnston,
  Kainulainen, \& Linz}]{Ragan_2014}
Ragan, S.~E., Henning, T., Tackenberg, J., {et~al.} 2014, Astronomy {\&}
  Astrophysics, 568, A73, \dodoi{10.1051/0004-6361/201423401}

\bibitem[{Reid {et~al.}(2016)Reid, Dame, Menten, \& Brunthaler}]{Reid_2016}
Reid, M.~J., Dame, T.~M., Menten, K.~M., \& Brunthaler, A. 2016, {ApJ}, 823,
  77, \dodoi{10.3847/0004-637x/823/2/77}

\bibitem[{{Rice} {et~al.}(2016){Rice}, {Goodman}, {Bergin}, {Beaumont}, \&
  {Dame}}]{Rice_2016}
{Rice}, T.~S., {Goodman}, A.~A., {Bergin}, E.~A., {Beaumont}, C., \& {Dame},
  T.~M. 2016, \apj, 822, 52, \dodoi{10.3847/0004-637X/822/1/52}

\bibitem[{{Roberts}(1969)}]{Roberts_1969}
{Roberts}, W.~W. 1969, \apj, 158, 123, \dodoi{10.1086/150177}

\bibitem[{{Robin} {et~al.}(2003){Robin}, {Reyl{\'e}}, {Derri{\`e}re}, \&
  {Picaud}}]{Robin_2003}
{Robin}, A.~C., {Reyl{\'e}}, C., {Derri{\`e}re}, S., \& {Picaud}, S. 2003,
  \aap, 409, 523, \dodoi{10.1051/0004-6361:20031117}

\bibitem[{Robitaille {et~al.}(2017)Robitaille, Beaumont, Qian, Borkin, \&
  Goodman}]{glueviz_2017}
Robitaille, T., Beaumont, C., Qian, P., Borkin, M., \& Goodman, A. 2017,
  Glueviz V0.13.1: Multidimensional Data Exploration, 0.13.1, Zenodo,
  \dodoi{10.5281/zenodo.1237692}.
\newblock \url{https://zenodo.org/record/1237692}

\bibitem[{{Rosolowsky} {et~al.}(2008){Rosolowsky}, {Pineda}, {Kauffmann}, \&
  {Goodman}}]{Rosolowsky_2008}
{Rosolowsky}, E.~W., {Pineda}, J.~E., {Kauffmann}, J., \& {Goodman}, A.~A.
  2008, \apj, 679, 1338, \dodoi{10.1086/587685}

\bibitem[{{Saito} {et~al.}(2012){Saito}, {Hempel}, {Minniti}, {Lucas},
  {Rejkuba}, {Toledo}, {Gonzalez}, {Alonso-Garc{\'\i}a}, {Irwin},
  {Gonzalez-Solares}, {Hodgkin}, {Lewis}, {Cross}, {Ivanov}, {Kerins},
  {Emerson}, {Soto}, {Am{\^o}res}, {Gurovich}, {D{\'e}k{\'a}ny}, {Angeloni},
  {Beamin}, {Catelan}, {Padilla}, {Zoccali}, {Pietrukowicz}, {Moni Bidin},
  {Mauro}, {Geisler}, {Folkes}, {Sale}, {Borissova}, {Kurtev}, {Ahumada},
  {Alonso}, {Adamson}, {Arias}, {Band yopadhyay}, {Barb{\'a}}, {Barbuy},
  {Baume}, {Bedin}, {Bellini}, {Benjamin}, {Bica}, {Bonatto}, {Bronfman},
  {Carraro}, {Chen{\`e}}, {Clari{\'a}}, {Clarke}, {Contreras}, {Corvill{\'o}n},
  {de Grijs}, {Dias}, {Drew}, {Fari{\~n}a}, {Feinstein},
  {Fern{\'a}ndez-Laj{\'u}s}, {Gamen}, {Gieren}, {Goldman},
  {Gonz{\'a}lez-Fern{\'a}ndez}, {Grand }, {Gunthardt}, {Hambly}, {Hanson},
  {He{\l}miniak}, {Hoare}, {Huckvale}, {Jord{\'a}n}, {Kinemuchi}, {Longmore},
  {L{\'o}pez-Corredoira}, {Maccarone}, {Majaess}, {Mart{\'\i}n}, {Masetti},
  {Mennickent}, {Mirabel}, {Monaco}, {Morelli}, {Motta}, {Palma}, {Parisi},
  {Parker}, {Pe{\~n}aloza}, {Pietrzy{\'n}ski}, {Pignata}, {Popescu}, {Read},
  {Rojas}, {Roman-Lopes}, {Ruiz}, {Saviane}, {Schreiber}, {Schr{\"o}der},
  {Sharma}, {Smith}, {Sodr{\'e}}, {Stead}, {Stephens}, {Tamura}, {Tappert},
  {Thompson}, {Valenti}, {Vanzi}, {Walton}, {Weidmann}, \&
  {Zijlstra}}]{Saito_2012}
{Saito}, R.~K., {Hempel}, M., {Minniti}, D., {et~al.} 2012, \aap, 537, A107,
  \dodoi{10.1051/0004-6361/201118407}

\bibitem[{{Shetty} \& {Ostriker}(2006)}]{Shetty_2006}
{Shetty}, R., \& {Ostriker}, E.~C. 2006, \apj, 647, 997, \dodoi{10.1086/505594}

\bibitem[{Smith {et~al.}(2014)Smith, Glover, Clark, Klessen, \&
  Springel}]{Smith_2014}
Smith, R.~J., Glover, S. C.~O., Clark, P.~C., Klessen, R.~S., \& Springel, V.
  2014, Monthly Notices of the Royal Astronomical Society, 441, 1628,
  \dodoi{10.1093/mnras/stu616}

\bibitem[{{Sousbie} {et~al.}(2011){Sousbie}, {Pichon}, \&
  {Kawahara}}]{Sousbie_2011}
{Sousbie}, T., {Pichon}, C., \& {Kawahara}, H. 2011, \mnras, 414, 384,
  \dodoi{10.1111/j.1365-2966.2011.18395.x}

\bibitem[{{Springel}(2010)}]{Springel_2010}
{Springel}, V. 2010, \mnras, 401, 791, \dodoi{10.1111/j.1365-2966.2009.15715.x}

\bibitem[{{Tackenberg} {et~al.}(2013){Tackenberg}, {Beuther}, {Plume},
  {Henning}, {Stil}, {Walmsley}, {Schuller}, \& {Schmiedeke}}]{Tackenberg_2013}
{Tackenberg}, J., {Beuther}, H., {Plume}, R., {et~al.} 2013, \aap, 550, A116,
  \dodoi{10.1051/0004-6361/201220140}

\bibitem[{{Turk} {et~al.}(2011){Turk}, {Smith}, {Oishi}, {Skory}, {Skillman},
  {Abel}, \& {Norman}}]{Turk_2011}
{Turk}, M.~J., {Smith}, B.~D., {Oishi}, J.~S., {et~al.} 2011, \apjs, 192, 9,
  \dodoi{10.1088/0067-0049/192/1/9}

\bibitem[{Vall\'{e}e(2008)}]{Vallee_2008}
Vall\'{e}e, J.~P. 2008, {AJ}, 135, 1301, \dodoi{10.1088/0004-6256/135/4/1301}

\bibitem[{{van der Walt} {et~al.}(2011){van der Walt}, {Colbert}, \&
  {Varoquaux}}]{numpy}
{van der Walt}, S., {Colbert}, S.~C., \& {Varoquaux}, G. 2011, Computing in
  Science and Engineering, 13, 22, \dodoi{10.1109/MCSE.2011.37}

\bibitem[{Wang {et~al.}(2016)Wang, Testi, Burkert, Walmsley, Beuther, \&
  Henning}]{Wang_2016}
Wang, K., Testi, L., Burkert, A., {et~al.} 2016, The Astrophysical Journal
  Supplement Series, 226, 9, \dodoi{10.3847/0067-0049/226/1/9}

\bibitem[{Wang {et~al.}(2015)Wang, Testi, Ginsburg, Walmsley, Molinari, \&
  Schisano}]{Wang_2015}
Wang, K., Testi, L., Ginsburg, A., {et~al.} 2015, Monthly Notices of the Royal
  Astronomical Society, 450, 4043, \dodoi{10.1093/mnras/stv735}

\bibitem[{{Wang} {et~al.}(2014){Wang}, {Zhang}, {Testi}, {van der Tak}, {Wu},
  {Zhang}, {Pillai}, {Wyrowski}, {Carey}, {Ragan}, \& {Henning}}]{Wang_2014}
{Wang}, K., {Zhang}, Q., {Testi}, L., {et~al.} 2014, \mnras, 439, 3275,
  \dodoi{10.1093/mnras/stu127}

\bibitem[{{Wilson}(1999)}]{Wilson_1999}
{Wilson}, T.~L. 1999, Reports on Progress in Physics, 62, 143,
  \dodoi{10.1088/0034-4885/62/2/002}

\bibitem[{{Zhang} {et~al.}(2019){Zhang}, {Kainulainen}, {Mattern}, {Fang}, \&
  {Henning}}]{Zhang_2019}
{Zhang}, M., {Kainulainen}, J., {Mattern}, M., {Fang}, M., \& {Henning}, T.
  2019, \aap, 622, A52, \dodoi{10.1051/0004-6361/201732400}

\bibitem[{Zucker {et~al.}(2015)Zucker, Battersby, \& Goodman}]{Zucker_2015}
Zucker, C., Battersby, C., \& Goodman, A. 2015, {ApJ}, 815, 23,
  \dodoi{10.1088/0004-637x/815/1/23}

\bibitem[{{Zucker} {et~al.}(2018){Zucker}, {Battersby}, \&
  {Goodman}}]{Zucker_2018a}
{Zucker}, C., {Battersby}, C., \& {Goodman}, A. 2018, \apj, 864, 153,
  \dodoi{10.3847/1538-4357/aacc66}

\bibitem[{{Zucker} \& {Chen}(2018)}]{Zucker_2018b}
{Zucker}, C., \& {Chen}, H. H.-H. 2018, \apj, 864, 152,
  \dodoi{10.3847/1538-4357/aad3b5}

\bibitem[{{Zucker} {et~al.}(2019){Zucker}, {Speagle}, {Schlafly}, {Green},
  {Finkbeiner}, {Goodman}, \& {Alves}}]{Zucker_2019}
{Zucker}, C., {Speagle}, J.~S., {Schlafly}, E.~F., {et~al.} 2019, \apj, 879,
  125, \dodoi{10.3847/1538-4357/ab2388}

\end{thebibliography}
\end{document}